\documentclass[9pt]{extarticle}
\usepackage{amsmath,amssymb,amsthm,bm}
\usepackage{mathtools}
\usepackage{geometry}
\usepackage{authblk}
\usepackage{graphicx}
\usepackage{subcaption}
\usepackage{xcolor}
\usepackage{hyperref}
\usepackage{siunitx}
\usepackage[version=4]{mhchem}
\usepackage{algorithm}
\usepackage{algpseudocode}
\usepackage{booktabs}
\usepackage[backend=bibtex,maxnames=99,minnames=1,doi=true,url=false,eprint=false,citestyle=numeric-comp,sorting=none]{biblatex}

\newcommand{\rcut}{r_\text{cut}}
\newcommand{\Rcut}{\mathcal{R}}
 
\newcolumntype{H}{>{\setbox0=\hbox\bgroup}c<{\egroup}@{}}

\begin{document}

\title{Flexible Cutoff Learning: Optimizing Machine Learning Potentials After Training}

\author[a,b]{Rick Oerder}
\author[b]{Jan Hamaekers}
\affil[a]{{\small Institute for Numerical Simulation, University of Bonn, Friedrich-Hirzebruch-Allee 7, 53115 Bonn, Germany}}
\affil[b]{{\small Fraunhofer Institute for Algorithms and Scientific Computing SCAI, Schloss Birlinghoven 1, \mbox{53757 Sankt Augustin}, Germany}}

\date{\today}
\maketitle

\begin{abstract}
    We introduce Flexible Cutoff Learning (FCL), a method for training machine learning interatomic potentials (MLIPs) whose cutoff radii can be adjusted after training. Unlike conventional MLIPs that fix the cutoff radius during training, FCL models are trained by randomly sampling cutoff radii independently for each atom. The resulting model can then be deployed with different per-atom cutoff radii depending on the application, enabling application-specific optimization of the accuracy-cost tradeoff. Using a differentiable cost model, these per-atom cutoffs can be optimized for specific target systems after training. We demonstrate FCL with a modified MACE architecture trained on the MAD dataset. For a subset featuring molecular crystals, optimized per-atom cutoffs reduce computational cost by more than 60\% while increasing force errors by less than 1\%. These results show that FCL enables training of a single general-purpose MLIP that can be adapted to diverse applications through post-training cutoff optimization, eliminating the need for retraining.
\end{abstract}

\section{Introduction}

Machine learning interatomic potentials (MLIPs) have become an essential tool in computational materials science and chemistry. When trained on large datasets generated with first-principles methods such as density functional theory, MLIPs provide accurate surrogate models of the Born-Oppenheimer potential energy surface while scaling linearly with system size. Supported by diverse datasets \cite{Eastman2023, deng_2023_chgnet, Schmidt2024, barroso_omat24, levine2025omol25, Mazitov2025, levine2025opoly26}, recent efforts have focused on developing foundational models \cite{batatia2025foundation, kovacs2025maceoff, wood2025, kim_optimizing_2025} that cover large regions of chemical space relevant to materials physics and drug design.

Foundational MLIPs achieve linear scaling by introducing a cutoff radius that limits the maximum number of interactions per atom. Often, the message passing approach \cite{gilmer2017} is then used to extend the models' receptive field beyond a single cutoff sphere. For example, MACE \cite{batatia2022mace}, SevenNet \cite{kim_sevennet_mf_2024}, Orb \cite{neumann2024orbfastscalableneural}, and UMA \cite{wood2025} construct messages by aggregating information from all pairs within the cutoff sphere. Other architectures (e.g., DimeNet++ \cite{gasteiger_dimenet_2020, gasteiger_dimenetpp_2020}, Mattersim \cite{mattersim2024arxiv}, M3GNet \cite{chen2022m3gnet}, CHGNet \cite{deng_2023_chgnet}) rely on explicit three-body terms or higher-order interactions \cite{gasteiger2021gemnet, wang2023}.

In periodic systems, the number of pairwise terms computed by an MLIP with explicit two-body interactions grows as $\mathcal{O}(\rcut^3)$, and for three-body MLIPs, computational cost scales as $\mathcal{O}(\rcut^6)$. Consequently, the cutoff radius significantly influences computational effort. Typical values range from $\SI{4.0}{}$ to $\SI{6.0}{\AA}$, with models computing explicit three-body interactions often using separate cutoff radii for two-body and three-body operations (\autoref{tab:cutoff_radii}).

\begin{table}[ht]
\centering
\caption{Cutoff radii and the highest explicit interaction order $k$ for selected MLIPs. Values for the cutoff radius in brackets for M3GNet and CHGNet correspond to a separate cutoff radius for three-body interactions.}
\begin{tabular}{l c c c}
\hline
Model & {$\rcut$ / \si{\angstrom}} & $k$ & Reference \\
\hline
MACE-MP-0                & 6.0         & 2 & \cite{batatia2025foundation}    \\
MACE-OFF23-S             & 4.5         & 2 & \cite{kovacs2025maceoff}     \\
MACE-OFF23-M/L           & 5.0         & 2 & \cite{kovacs2025maceoff}     \\
MACE-OFF24-M             & 6.0         & 2 & \cite{kovacs2025maceoff}       \\
SevenNet-Omni            & 6.0         & 2 & \cite{kim_sevennet_mf_2024}    \\
UMA                      & 6.0         & 2 & \cite{wood2025}                \\
\hline
M3GNet                   & 5.0 (4.0)   & 3 & \cite{chen2022m3gnet}          \\
Mattersim                & 5.0 (4.0)   & 3 & \cite{mattersim2024arxiv}      \\
CHGNet                   & 5.0 (3.0)   & 3 & \cite{deng_2023_chgnet}        \\
\hline
\end{tabular}
\label{tab:cutoff_radii}
\end{table}

As suggested by these considerations, the cutoff radius is critical for balancing model accuracy and computational performance, yet it is treated as a static hyperparameter in all current foundational MLIPs. Once a model is trained, the cutoff radius is fixed and cannot be adjusted without retraining which may be prohibitively expensive for large models or datasets. Because foundational models must remain reliable across diverse systems, practitioners typically choose conservatively large cutoffs (e.g., $\SI{6}{\AA}$) to avoid systematically neglecting relevant interactions, even though many specialized applications could achieve comparable accuracy with smaller cutoffs and significantly reduced computational cost.

We propose \textit{Flexible Cutoff Learning} (FCL) to address this limitation. FCL conditions the model on the cutoff radius during training. By explicitly providing cutoff values as inputs, the model learns to adapt its predictions accordingly. This enables post-training optimization of the accuracy-cost tradeoff without requiring retraining. 

Our key contributions are the following:

\begin{itemize}
    \item \textbf{Post-training flexibility:} We promote the cutoff radius from a static hyperparameter to a dynamic variable that can be adjusted after training, enabling application-specific optimization without retraining.
    \item \textbf{Per-atom cutoffs:} Rather than using a single global cutoff radius, FCL allows each atom to have its own flexible cutoff, enabling fine-grained control over the accuracy-cost tradeoff.
    \item \textbf{Training methodology:} We introduce a training workflow that randomly samples cutoff radii during training, producing models that maintain smoothness and accuracy across different cutoff configurations.
    \item \textbf{Systematic optimization:} We demonstrate gradient-based optimization of per-atom cutoffs using a differentiable cost model, enabling systematic tuning for target systems.
\end{itemize}

We demonstrate FCL by training a modified MACE architecture on the MAD dataset, showing that optimal cutoffs can reduce computational cost by a substantial amount while force errors increase only slightly. FCL-trained models provide greater flexibility for adapting foundational MLIPs to specific applications, complementing existing finetuning workflows that optimize accuracy alone rather than the error-cost balance.

\paragraph*{Related Work} The Point Edge Transformer (PET) architecture \cite{pozdnyakov2023} employs a post-training protocol that may appear similar to our approach but differs fundamentally in flexibility and scope. PET uses \textit{Equivariant Coordinate System Ensemble} (ECSE), which symmetrizes predictions of a non-equivariant backbone model by averaging over local coordinate frames constructed from atom pairs within a cutoff sphere. To make this computationally feasible, ECSE introduces a second \textit{inner cutoff radius} that determines which pairs form coordinate frames in the post-training symmetrization step.

Critically, PET's inner cutoff selection follows a fixed, environment-dependent heuristic rather than being freely optimizable. The backbone model itself is trained with a static, global cutoff radius, and the inner cutoff adjusts only the coordinate frame selection during post-processing. In contrast, FCL trains the neural network to handle arbitrary cutoff configurations, enabling full optimization of per-atom cutoffs for specific applications.

Recently, Han et al. \cite{han2026} introduced the concept of a \textit{smooth dynamic cutoff} protocol that determines the cutoff radius in a way that approximately reproduces a fixed target value for the average number of neighbors of each atom. During and after training, the number of neighbors is kept fixed. In contrast, FCL models learn to adapt their predictions to any valid cutoff configuration (even after training) and not only the ones that yield a constant neighbor count. 
In essence, FCL enables post-training optimization of the accuracy-cost tradeoff for specific target systems, allowing cutoff radii to be determined by application requirements rather than predefined rules.

\paragraph*{}
The remainder of this paper is organized as follows. In \autoref{sec:methods} we introduce the core ideas of FCL and discuss an approach for optimizing the accuracy-cost relationship of FCL models. Results obtained by training an FCL-adapted variant of the MACE architecture on the MAD dataset and the optimization of per-element cutoff radii on subsets are presented \autoref{sec:Results}. Finally, the results and limitations of this work are discussed in \autoref{sec:discussion} before coming to an conclusion in \autoref{sec:conclusion}.

\section{Methods}
\label{sec:methods}
\subsection{Preliminaries}

\paragraph*{Neighborhood truncation} Limiting interactions to a finite cutoff radius $\rcut$ bounds the number of neighbors that contribute to each atomic environment. This is crucial for computational efficiency (see below), as it enables efficient neighbor list construction, domain decomposition, and scalability to large systems and many compute devices.
In molecular dynamics simulations, atoms may cross the cutoff boundary and leave or enter the cutoff sphere of an atom. To avoid discontinuities in that case, taper  functions (sometimes called shifting functions or cutoff functions) $s(r)$ are used to suppress interactions as the interatomic distance $r$ approaches $\rcut$ in way that preserves continuity. Typically, $s(\rcut)=0$ is required together with at least a vanishing first derivative at $r=\rcut$ so that both the potential and the forces are smooth. 

A common choice to achieve this is given by polynomials  such as

\begin{equation}
\label{eq:polynomial_cutoff}
s(r) = \left[ 1 - \frac{(p + 1)(p + 2)}{2} \left(\frac{r}{\rcut}\right)^{p}
+ p (p + 2) \left(\frac{r}{\rcut}\right)^{p + 1}
- \frac{p (p + 1)}{2} \left(\frac{r}{\rcut}\right)^{p + 2} \right] [r\leq \rcut]
\end{equation}

for $p >0$, where $[\cdot]$ denotes the Iversion bracket \cite{gasteiger_dimenet_2020}.

In MLIPs, these taper functions are usually applied directly to the outputs of neural network layers that model pairwise interactions, ensuring that the learned potential strictly respects the cutoff and remains differentiable.

\paragraph*{Computational Complexity} \label{sec:computational_complexity}
The computational implications of the cutoff radius differ between finite molecules and periodic materials. For a molecule with $n$ atoms and no periodic boundary conditions, the number of neighbors of any atom is at most $n-1$, even as $\rcut \to \infty$. Consequently, explicit 2-body interactions is as most $\mathcal{O}(n^2)$, and the number of higher-body terms is bounded by the finite system size as well.

In periodic systems, by contrast, the number of neighbors increases with $\rcut$ without a global upper bound. Assuming a homogeneous and isotropic atomic distribution with number density $\rho$, the average number of neighbors within a sphere of radius $\rcut$ around an atom scales as

\begin{equation}
    z(\rcut) \propto \rho\,\rcut^3.
\end{equation}

For an explicit $k$-body term centered on atom $i$, one chooses $(k-1)$ neighbors among these $z(\rcut)$ atoms, so the number of such interaction terms per atom scales as

\begin{equation}
    \binom{z(\rcut)}{k-1}
= \mathcal{O}\!\big(z(\rcut)^{\,k-1}\big)
= \mathcal{O}\!\big(\rcut^{3(k-1)}\big).
\end{equation}

Thus, explicit 2-body, 3-body, and 4-body contributions scale as $\mathcal{O}(\rcut^3)$, $\mathcal{O}(\rcut^6)$, and $\mathcal{O}(\rcut^9)$ per atom, respectively. 

As discussed previously, most MLIPs avoid explicit many-body summations beyond low order (compare \autoref{tab:cutoff_radii}). In all cases, the cutoff radius determines the size of the local environment and thus influences the computational effort and the attainable accuracy in a significant way.

\subsection{Flexible Cutoff Learning}
\label{subsec:flexible_cutoff_learning}

\begin{figure}
    \centering
    \includegraphics[width=0.95\linewidth]{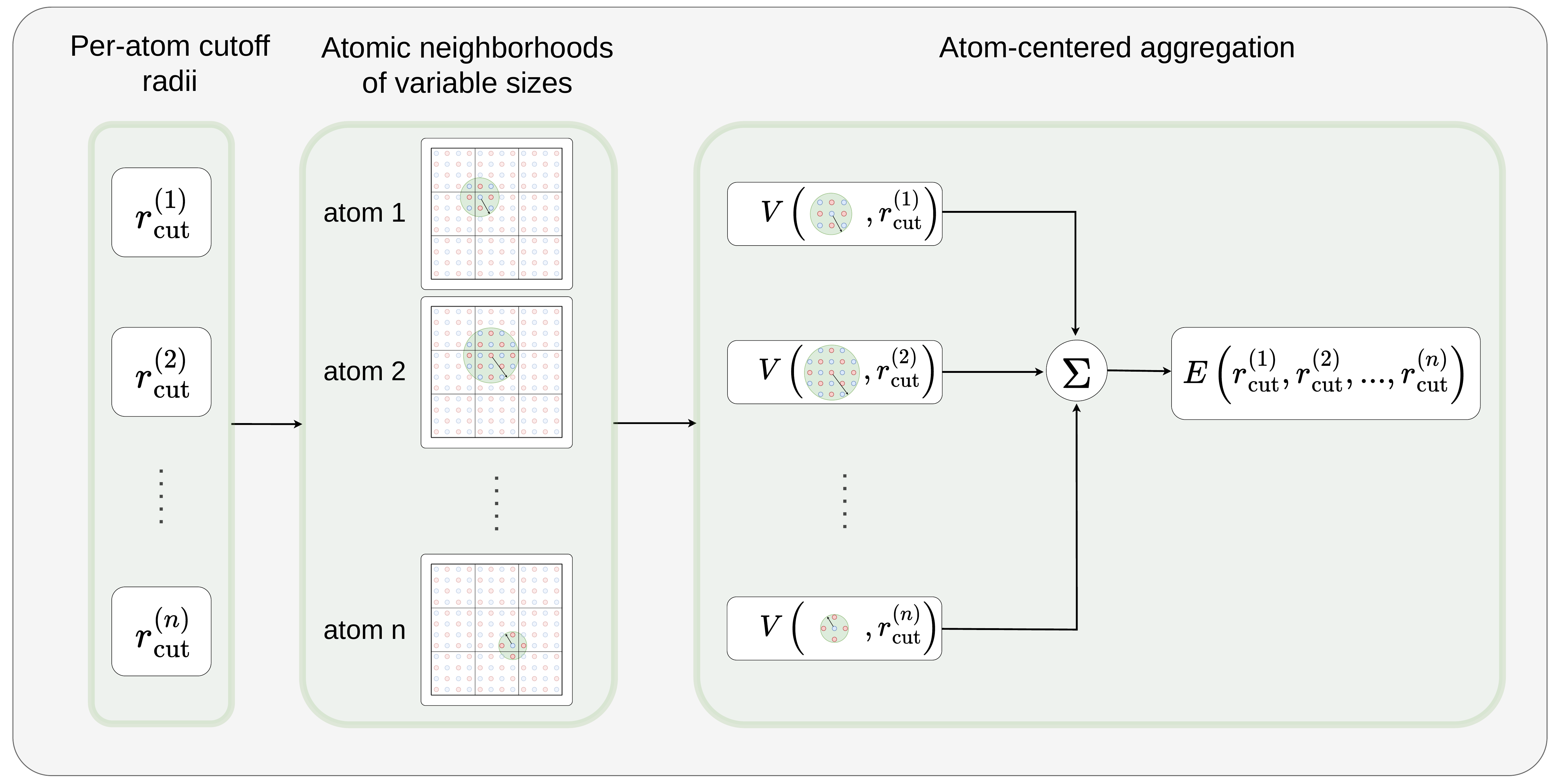}
    \caption{Simplified representation of a model that depends explicitly on atom-wise cutoff radii. For a fixed geometry, the model can be evaluated for different realizations of $\rcut^{(1)},...,\rcut^{(n)}$. In this schematic figure, the model is assumed to be decomposable into on-site contributions that are computed from atom-centered atomic environments. In particular, each contribution is obtained by evaluating a function (denoted as $V$) on atomic environments of different sizes that depend on the choice of the atom-wise cutoff radii.}
    \label{fig:model}
\end{figure}

The main goal of Flexible Cutoff Learning is to obtain a model for which the cutoff radii can be changed even after training has been finished.  
To that end, instead of using a fixed cutoff radius, cutoff radii are sampled stochastically during training to learn a model that can adjust its prediction to a variety of cutoff radii. 
In what follows, we discuss our approach of adapting conventional MLIPs $f_\theta(x_1, x_2, \ldots, x_n)$, where $x_i = (\mathbf{r}_i, Z_i)$ are tuples of Cartesian nuclear coordinates $\mathbf{r}_i$ and $Z_i$ encodes the element type, to FCL. Instead of using a global and static cutoff radius, we modify existing architectures to have an explicit dependency on the per-atom cutoff radii  $\rcut^{(i)}$, which enables us to adjust the neighbor list sparsity for each atom individually. These modified models $f_\theta(x_1, x_2, \ldots, x_n;  \rcut^{(1)},\rcut^{(2)},...,\rcut^{(n)})$ operate on graph topologies in which the set of neighbors of an atom is given by 

\begin{equation}
    \mathcal{N}_i = \{j \in \{1,\ldots,n\} \text{ s.t. } \lVert \mathbf{r}_i - \mathbf{r}_j \rVert_2 \leq \mu(\rcut^{(i)}, \rcut^{(j)}) \}
    \label{eq:mixing_rule}
\end{equation}

where $\mu$ is a mixing rule. If not stated otherwise, $\mu$ is the arithmetic mean throughout the remainder of this work. Constructing neighbor lists according to equation \eqref{eq:mixing_rule} enables precise control over the computational cost and allows the graph topology to be adapted to individual regions of the simulation domain.
For notational convenience, we write 

\begin{equation}
    f_\theta(\mathcal{X};\Rcut) \equiv f_\theta(x_1, x_2, \ldots, x_n;  \rcut^{(1)},\rcut^{(2)},...,\rcut^{(n)})
\end{equation} 

by introducing $\mathcal{X}=\{x_1, x_2, \ldots, x_n\}$, $\Rcut=\{\rcut^{(1)},\rcut^{(2)},...,\rcut^{(n)}\}$. 

\paragraph*{Architecture changes} Varying the set of cutoff radii $\rcut^{(1)},\rcut^{(2)}, \ldots, \rcut^{(n)}$ for a fixed atom geometry during and after training requires a few but easy-to-implement changes of existing architectures to ensure the differentiability of predictions. In the following, we will present a high-level overview on changes that are necessary to adapt a specific architecture to the FCL approach.

\begin{enumerate}
    \item \textbf{Computation of the taper function} To account for the local and dynamic nature of the cutoff radius in our framework, it is necessary to compute the cutoff radius for each pair of adjacent atoms according to the mixing rule \eqref{eq:mixing_rule} at each forward pass as $m_{ij} = \mu(\rcut^{(i)}, \rcut^{(j)})$.  
    The resulting value should be provided to the method that computes the taper function as an explicit argument. That is, the taper function should be treated as a bivariate function $s(r_{ij}, m_{ij})$ rather than a function $s(r_{ij})$. This change ensures differentiable predictions for all cutoff radii and for most architectures, it requires modification of only a few lines of code.
    \item \textbf{Conditioning on cutoff radii} To enable the model to adapt its predictions based on the cutoff configuration, it must be conditioned on the cutoff radius of each atom. For most MLIPs, this can be achieved by employing a trainable scalar embedding function $e: \mathbb{R} \to \mathbb{R}^d$ that maps each cutoff value $\rcut^{(i)}$ to a $d$-dimensional feature vector, which is then added to the initial node features for atom $i$. This allows the model to learn different representations for the same atomic environment when observed at different cutoff radii. The embedding function can be implemented as a feedforward network. Other relevant parts of the model, such as edge features or readout layers can be informed in similar ways. 
\end{enumerate}

In \autoref{subsec:FCL-mace} we describe how the MACE architecture can be modified as described above to represent different cutoff configurations while remaining differentiable at each induced cutoff boundary.
A simplified representation of a model that explicitly depends on per-atom cutoff radii is depicted in \autoref{fig:model}. As discussed in \autoref{subsec:optimization}, we are interested in optimizing the configuration of cutoff radii. To that end, it is important that model predictions, such as energy or forces, are differentiable with respect to $\Rcut$. All changes made to existing architectures should respect this requirement. 

\begin{figure}[htbp]
  \centering
  \begin{subfigure}[b]{0.48\textwidth}
    \centering
    \includegraphics[width=0.7\linewidth]{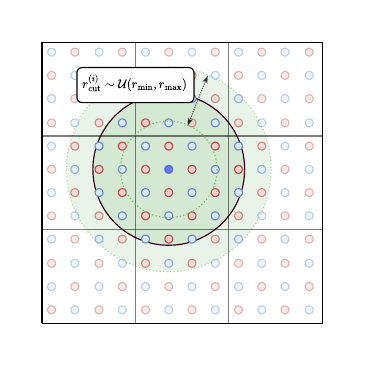}
    \caption{}
    \label{fig:schematic_sampling}
  \end{subfigure}
  \hfill
  \begin{subfigure}[b]{0.48\textwidth}
    \centering
    \includegraphics[width=0.98\linewidth]{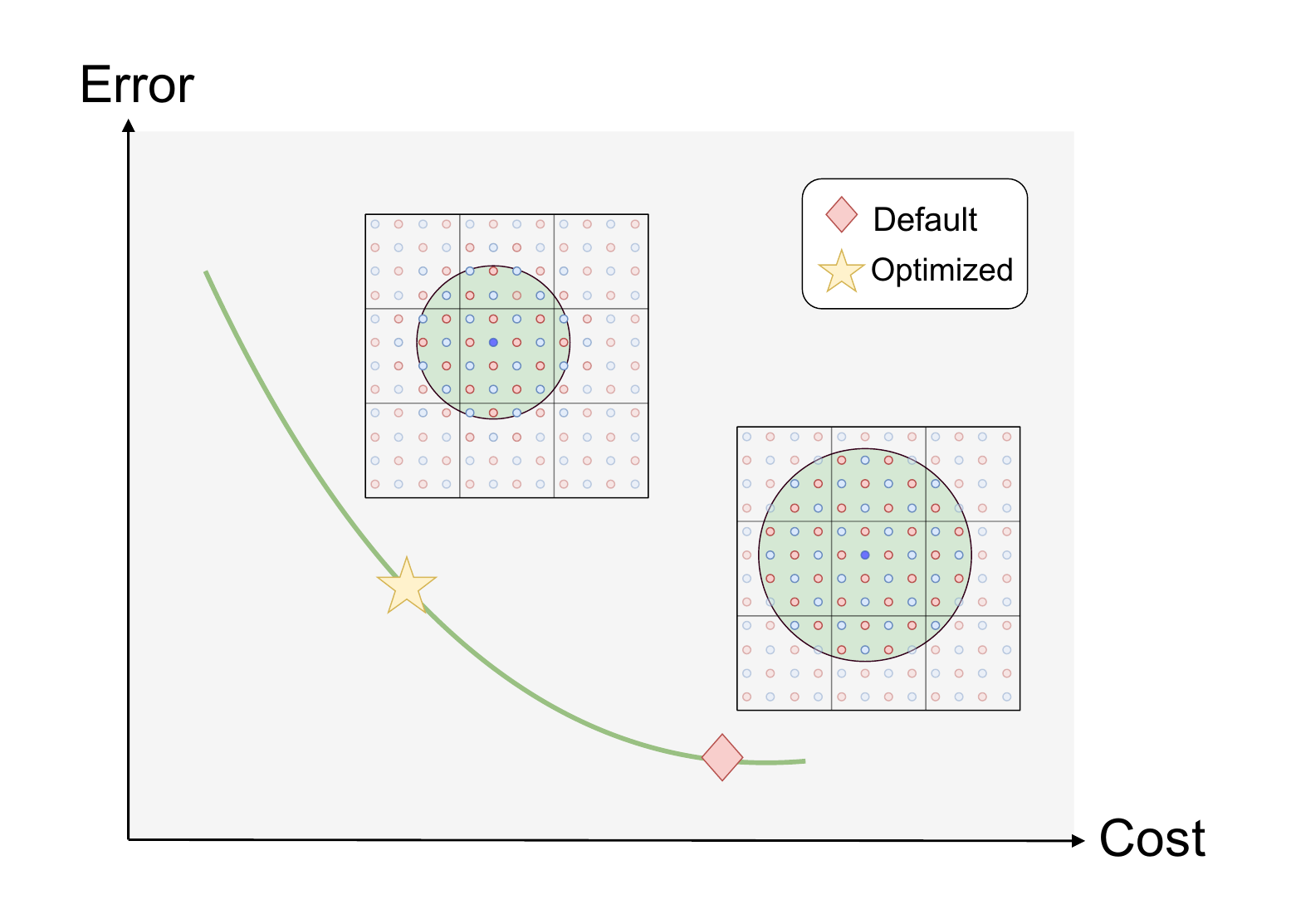}
    \caption{}
    \label{fig:schematic_optimization}
  \end{subfigure}
  \label{fig:FCL-During-After-Training}
  \caption{Schematic overview of the Flexible Cutoff Learning and the optimization of cutoff radii after training. (a) Per-atom cutoff radii $\rcut^{(i)}$ are sampled during the training process. (b) Optimization of error and cost with respect to cutoff radii is performed after training. Smaller receptive fields lead to faster evaluation but may reduce the model accuracy.}
\end{figure}

\paragraph*{Training} The goal of FCL is to train a model that generalizes well to a variety of different cutoff configurations for maximum flexibility after training. Therefore, it is necessary to train the model with suitable configurations representative of cutoff radii relevant for downstream applications. To avoid the immense cost of constructing a full grid of collocation points for each data point, we perform random sampling of per-atom cutoff radii from a uniform distribution $\mathcal{U}(r_\text{min}, r_\text{max})$ at each training step (see \autoref{fig:schematic_sampling}). Stochastic sampling for conditioning neural networks on relevant context was demonstrated previously in \cite{Dobberstein2024}. Each data point in a mini-batch $\mathcal{M}$ of size $B$ is augmented by a set of atom-wise cutoff radii $\Rcut$ that are sampled randomly at each training step and a loss function

\begin{equation}
    \mathcal{L} \left( \theta \right) = \frac{1}{B} \sum_{(\mathcal{X},y,\Rcut) \in \mathcal{M}} \ell \left( f_\theta (\mathcal{X}; \Rcut), y \right)
\end{equation}

is optimized with respect to the model parameters where $\ell$ denotes a per-sample loss function. For simplicity, we omitted additional loss terms for matching of other properties (such as forces, partial charges or stress tensors) in our notation. 

The interval $[r_\text{min}, r_\text{max}]$ should be chosen in a way that allows flexible post-training adjustments, e.g. $r_\text{min}=\SI{3.5}{\AA}$ and $r_\text{min}=\SI{7.0}{\AA}$. In general, the specific choice of the interval will also depend on the number of message passing layers that influence the effective receptive field of a model.

Note that while the model depends on $\Rcut$, the neighbor list topology according to \eqref{eq:mixing_rule} is constructed before the forward pass and held fixed during backpropagation. Gradients flow through the taper function $s(r_{ij}, m_{ij})$ but not through the discrete neighbor selection. This ensures computational efficiency while maintaining differentiability with respect to atomic positions.

The pseudocode describing the details of the training loop can be found in Algorithm \autoref{pseudocode:training}. 

\begin{algorithm}
\caption{Pseudocode for the Flexible Cutoff Learning approach. }
\begin{algorithmic}[1]
\Require Dataset $D = \{(\mathcal{X}_i, y_i)\}_{i=1}^N$, batch size $B$, number of epochs $E$, initial parameters $\theta_0$, optimizer \texttt{OPT},  lower bound for sampling cutoff radii $r_\text{min}$, upper bound for sampling cutoff radii $r_\text{max}$
\State Initialize model parameters $\theta \gets \theta_0$
\For{epoch $= 1$ to $E$}
    \State Shuffle dataset $D$
    \For{each mini-batch $M$ of size $B$ in $D$}
        \State Initialize mini-batch loss: $L \gets 0$
        \For{$(\mathcal{X},y)$ in mini-batch $M$}
            \State Initialize empty set for cutoff radii: $\Rcut \gets \{ \}$
            \For{each node $i$ in $\mathcal{X}$}
                \State Sample $c_i \sim \mathcal{U}( r_\text{min}, r_\text{max})$
                \State Set $\Rcut[i] \gets c_i$
            \EndFor
            \State Update loss: $L \gets L+ \ell \left( f_\theta (\mathcal{X}; \Rcut), y \right)$
        \EndFor
        \State Compute gradients $g$: $g \gets \nabla_\theta L$
        \State Update parameters: $\theta \gets \texttt{OPT}(\theta, g)$
    \EndFor
\EndFor
\State \Return $\theta$
\end{algorithmic}
\label{pseudocode:training}
\end{algorithm}

\paragraph*{Evaluation} Once a model is trained, it can be used for inference, for example, in MD simulations or for geometry optimizations by specifying a set of static cutoff radii $\Rcut^\text{fixed} = \{\rcut^{(1),\text{fixed}},\rcut^{(2),\text{fixed}},\ldots, \rcut^{(n)}\}$ resulting in 

\begin{equation}
    f_\theta(\mathcal{X}) \equiv  f_\theta(\mathcal{X}; \Rcut^\text{fixed}).
\end{equation}

In general, there will be an unlimited number of realizations for $\Rcut^\text{fixed}$ each leading to a model with its own error-cost tradeoff.  The most trivial way to evaluate a model is by setting all atomic cutoff radii to the same value. As described in \autoref{subsec:optimization}, error-cost optimal realizations of the cutoff radii can be chosen through gradient based optimization, as sketched in \autoref{fig:schematic_optimization}. 

\subsection{Optimization of Cutoff Radii}
\label{subsec:optimization}

The computational cost of an MLIP evaluation increases with the number of pairwise interactions, which strongly depends on the cutoff radius (\autoref{sec:computational_complexity}). Larger cutoffs capture longer-range interactions but increase cost; smaller cutoffs reduce cost but may sacrifice accuracy. For a given target application and acceptable error tolerance, there exists an optimal cutoff configuration that balances these competing objectives.

For an FCL-trained model, an optimal configuration can be found by optimizing the accuracy-cost tradeoff on a calibration set $\mathcal{D}_\text{cal}$ representative of the target application while keeping the model weights fixed. We describe this optimization procedure below.

For simplicity, we focus on the optimization of the cutoff radii per-element $\Rcut_\text{E} \coloneq \{ a^{(\ce{H})}, a^{(\ce{He})}, a^{(\ce{Li})}, \ldots \}$. 
To that end, we define a map for broadcasting $\mathcal{R}_E$ to the atoms of an atomistic structure $\mathcal{X}=(\mathbf{r}_i, Z_i)$ as

\begin{equation}
    \kappa: \{ (\mathbf{r}_i, Z_i)\}_{i=1}^n\mapsto \{ a^{(Z_i)}\}_{i=1}^n .
\end{equation}

We define an error metric that depends on the cutoff radius chosen for each atom as 

\begin{equation}
    \epsilon (\Rcut_\text{E}) = \frac{1}{|\mathcal{D}_\text{cal}|} \sum_{(\mathcal{X},y) \in \mathcal{D}_\text{cal}} \ell \left( f_\theta \left(\mathcal{X}; \kappa(\mathcal{X}) \right), y \right),
\end{equation}

where $y$ are reference labels that are assumed to be consistent with the training data.

Next, we represent the computational cost in a differentiable fashion to allow for combined optimization of accuracy and performance. To that end, we consider the following differentiable model of the average computational cost per atom

\begin{equation}
    C(\Rcut_\text{E}) = \frac{1}{|\mathcal{D}_\text{cal}|} \sum_{i=1}^{|\mathcal{D}_\text{cal}|} \frac{1}{n_i} \sum_{j=1}^{n_i} {(a^{(Z_j)})}^{3},
\end{equation}

where $n_i$ denotes the number of atoms in the data point with index $i$. The assumption of this cost model is that the number density of atoms is constant across all structures in the calibration data set. In general, this will not strictly be the case and deviations of this assumption should be tested by evaluating the actual number of pairs per atom. We discuss the measurement of the computational cost in \autoref{sec:Results}.

Optimizing predictive accuracy and computational cost poses a multi-objective optimization scenario. We consider a scalarized target function that models the tradeoff between model accuracy and computational cost:

\begin{equation}
    T(\Rcut_\text{E}) =  \epsilon (\Rcut_\text{E}) + \lambda \cdot C(\Rcut_\text{E}).
    \label{eq:target_function_opt}
\end{equation}

The hyperparameter $\lambda$ controls the tradeoff between accuracy and computational cost. Different values of $\lambda$ yield different points on the Pareto frontier of this multi-objective optimization problem. In practice, $\lambda$ can be selected by sweeping over a range of values and choosing it based on acceptable error thresholds. 
In our experiments (\autoref{sec:Results}), we explore multiple values of $\lambda$ to characterize the accuracy-cost tradeoff in a comprehensive way.

For a fixed $\lambda$, optimal values for the element-specific cutoff radii at inference can then be found through a minimization of \eqref{eq:target_function_opt}:

\begin{equation}
    (a^{(\ce{H}),\ast}, a^{(\ce{He}),\ast}, a^{(\ce{Li}),\ast}, \ldots) = \underset{\Rcut_\text{E}}{\text{argmin }} T(\Rcut_\text{E})
    \label{eq:optimal_cutoff}
\end{equation}

Practically, optimization with respect to $\Rcut_\text{E}$ can be performed with gradient-based optimization techniques such as stochastic gradient descent or Adam \cite{kingma2014}, see \autoref{sec:Results}.

\section{Results}
\label{sec:Results}

\paragraph*{Flexible Cutoff Learning on the MAD dataset}
We evaluate Flexible Cutoff Learning on the Massive Atomic Diversity (MAD) dataset \cite{Mazitov2025}, which provides substantial chemical diversity in a lightweight format. We use the predefined 8:1:1 train/validation/test splits and employ the reported labels (energies and forces) without any post-processing.

Training follows a two-stage protocol. First, we pretrain a standard MACE model with a fixed global cutoff of $\SI{6}{\AA}$. Second, we perform Flexible Cutoff Learning (FCL) as described in \autoref{subsec:flexible_cutoff_learning} with $r_\text{min}=\SI{3.5}{\AA}$ and $r_\text{max}=\SI{7.0}{\AA}$, repeating this stage for three different seeds starting from the same pretrained checkpoint. Neighbor lists for each sample in the mini-batch are obtained by first computing a global neighbor list with the maximum per-atom cutoff and then pruning edges that exceed the pairwise cutoff inferred from the mixing rule. The results are averaged over the three seeds. We refer to the resulting models as FCL models. We describe the training workflow in detail below.

\begin{figure}[htbp]
  \centering
    \includegraphics[width=0.7\textwidth]{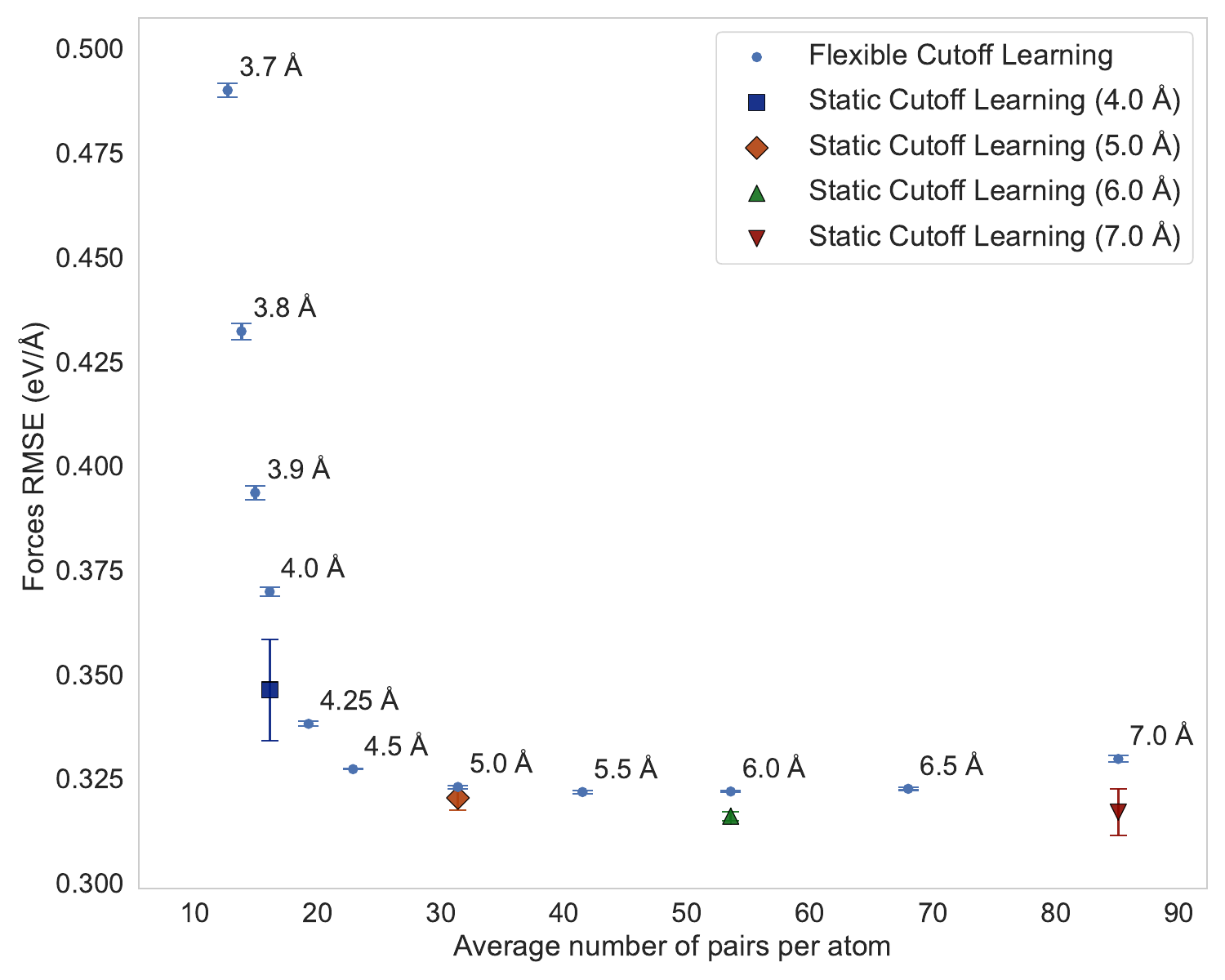}
  \caption{Test errors as a function of a global cutoff radius. Results for "Flexible Cutoff Learning" are obtained from evaluating a single model with different global cutoff radii. "Static Cutoff Learning" correspond to a separate model for each cutoff radius. All results are averaged over three training runs with different seeds. Error bars correspond to the empirical standard deviation.}
  \label{fig:Error-Cost-Curves-Test}
\end{figure}

To visualize performance across different cutoff configurations, we evaluate the FCL model on the test set using uniform cutoffs (i.e., $\rcut^{(1)} = \rcut^{(2)} = \ldots = \rcut^{(n)} = \rcut $). \autoref{fig:Error-Cost-Curves-Test} shows the root mean square error (RMSE) of force predictions and the average number of pairs per atom as a function of the global cutoff radius. The FCL model achieves an RMSE of approximately $\SI{0.370}{eV/\angstrom}$ at $\rcut = \SI{4.0}{\AA}$, which decreases to $\SI{0.325}{eV/\angstrom}$ at $\SI{5.0}{\AA}$ before plateauing. A slight increase in error is observed at $\SI{7.0}{\AA}$. Models trained with static cutoffs (also shown in \autoref{fig:Error-Cost-Curves-Test}) exhibit similar trends but achieve consistently lower errors, as expected: predicting accurately across a wide range of cutoffs is inherently more challenging than training specialized models for each cutoff value. 

The results demonstrate that small cutoff changes significantly impact both accuracy and cost. For example, increasing the cutoff from $\SI{3.7}{\AA}$ to $\SI{4.0}{\AA}$ raises the average number of pairs from 12.6 to 16.0, that is, a 28\% increase in computational cost. Because the MAD dataset contains systems with varying periodicity (3D, 2D, and 0D), the average number of pairs is lower than would be observed for purely 3D-periodic materials, where the $\mathcal{O}(\rcut^3)$ scaling has the strongest effect.

\begin{figure*}
  \begin{subfigure}{0.33\textwidth}
    \includegraphics[width=\linewidth]{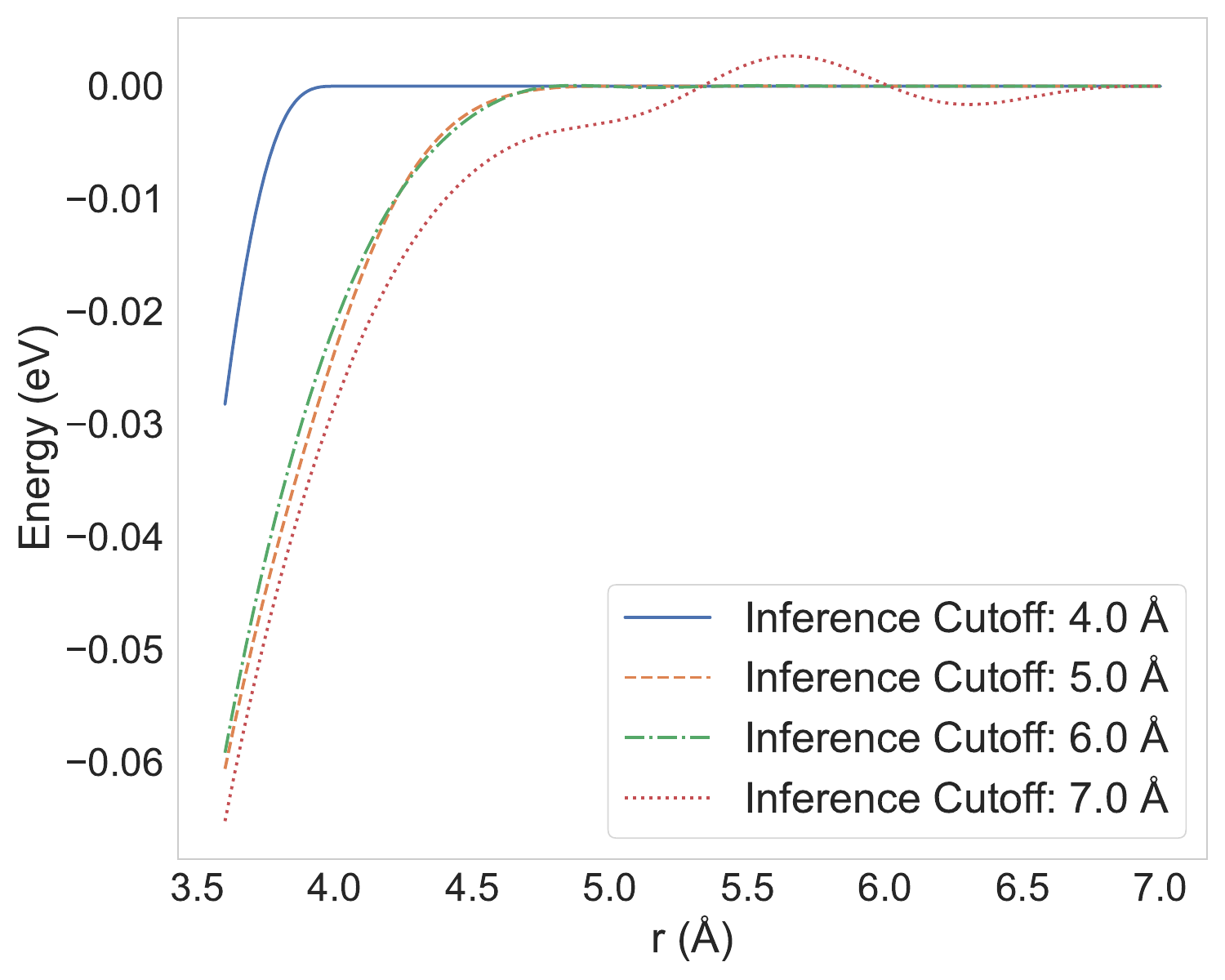}
    \caption{H-O}
  \end{subfigure}%
  \hfill
  \begin{subfigure}{0.33\textwidth}
     \includegraphics[width=\linewidth]{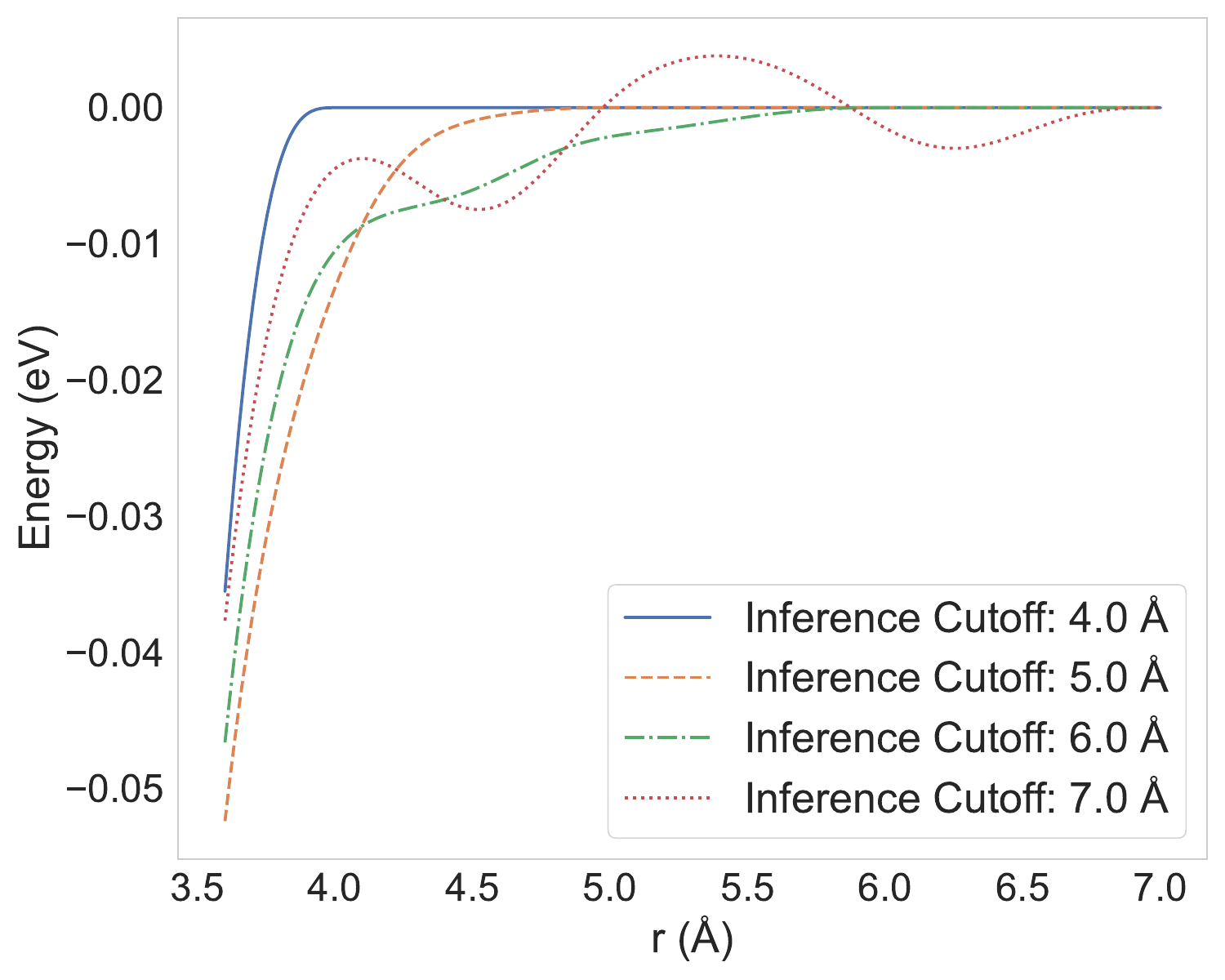}
    \caption{C-C}
  \end{subfigure}%
  \hfill
  \begin{subfigure}{0.33\textwidth}
    \includegraphics[width=\linewidth]{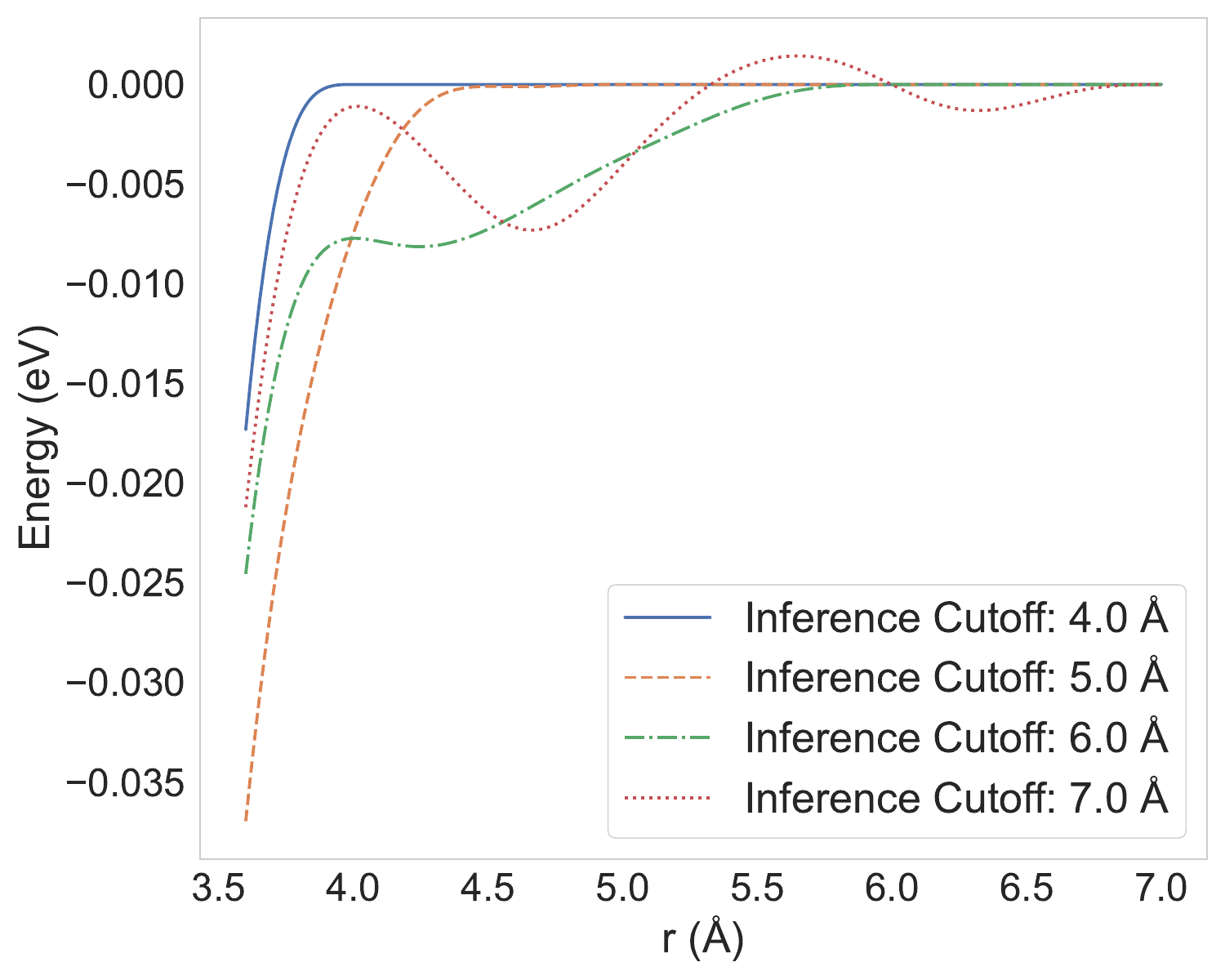}
    \caption{C-O}
  \end{subfigure}
  \caption{Energy as a function of distance in diatomic systems for different inference cutoff radii. All curves are obtained by evaluating a single FCL model.}
  \label{fig:pair_curves}
\end{figure*}

\paragraph*{Diatomic curves} 
To understand how the inference cutoff radius affects model predictions, we examine diatomic energy-distance curves obtained with the FCL model (\autoref{fig:pair_curves}) when evaluated with uniform cutoffs, i.e., $\rcut^{(1)}=\rcut^{(2)}=\rcut$ for $\rcut \in \{ \SI{4.0}{\AA}, \SI{5.0}{\AA}, \SI{6.0}{\AA}, \SI{7.0}{\AA} \}$. We focus on representative element pairs (H-O, C-C, C-O) that are frequent in the MAD dataset. The plots show the region near the cutoff radius where differences between inference cutoffs are most visible; full-range curves are provided in the appendix.

The curves for cutoff radii of $\SI{4.0}{\AA}$, $\SI{5.0}{\AA}$, and $\SI{6.0}{\AA}$ appear smooth and physically reasonable. However, curves evaluated with $\rcut = \SI{7.0}{\AA}$ exhibit oscillatory behavior in all three cases, indicating degraded performance at the boundary of the training distribution. During training, cutoff radii are sampled uniformly from $\mathcal{U}(r_\text{min}, r_\text{max})$ with $r_\text{max} = \SI{7.0}{\AA}$. While this cutoff value is included in training, the model has no training examples with larger cutoffs to constrain its behavior at the upper boundary. This makes it difficult for the model to learn smooth transitions at $r_\text{max}$ compared to interior values where the model observes both larger and smaller cutoffs during training. Without additional constraints such as hard boundary conditions, this could lead to reduced accuracy when evaluating at $r_\text{max}$. This could be one explanation for the increased error at $\SI{7.0}{\AA}$ observed in \autoref{fig:Error-Cost-Curves-Test}. In addition,  an increased empirical standard deviation over the three training runs for the static models with $\SI{7.0}{\AA}$ indicates that learning becomes less stable for larger cutoff radii, even when training with a static cutoff radius.  

Comparing curves across different cutoff radii reveals typical energy differences of approximately $\SI{0.01}{\eV}$ between adjacent cutoffs (e.g., $\SI{5.0}{\AA}$ vs. $\SI{6.0}{\AA}$). While individual pair contributions from increasing an atom's cutoff radius are small, the cubic scaling of neighbor count in periodic systems means these differences can accumulate to notable contributions to the total energy.

\paragraph*{Training details} 
In the first pretraining stage, we optimized an unmodified MACE model with a static cutoff of $\SI{6.0}{\AA}$ for 200 epochs. The loss function combined L1 terms for per-atom energies (weight $0.1$) and atomic forces (weight $1.0$); stress components were not included. We used the AdamW optimizer \cite{loshchilov2018} (learning rate $0.02$, weight decay $10^{-8}$) with a batch size of 50. A \texttt{MultiStepLR} scheduler reduced the learning rate by a factor of $0.5$ at epochs 10, 20, and 100. The MACE architecture used 2 interaction layers, 10 Bessel basis functions, 5 polynomial cutoff terms, correlation order 1, maximum angular momentum $\ell_{\text{max}} = 1$, and hidden dimension 32, resulting in approximately $373{,}000$ parameters.

For Flexible Cutoff Learning, we initialized the modified model from the pretrained weights (see \autoref{subsec:FCL-mace}). The modified radial embedding block was initialized from scratch, adding approximately $9{,}000$ parameters (less than 3\% increase). We retained the batch size of 50 and the same loss weights. The AdamW optimizer used a reduced learning rate of $0.001$ (weight decay $10^{-8}$). A \texttt{ReduceLROnPlateau} scheduler reduced the learning rate by $0.5$ after 20 epochs without improvement in the combined validation loss evaluated at cutoffs of $\SI{4.0}{\AA}$, $\SI{5.0}{\AA}$, and $\SI{6.0}{\AA}$, with a minimum learning rate of $10^{-5}$. Training proceeded for up to 500 epochs with early stopping (patience 50 epochs) based on the same validation metric.

For baseline comparisons, we trained unmodified MACE models with static cutoffs of $\SI{4.0}{\AA}$, $\SI{5.0}{\AA}$, $\SI{6.0}{\AA}$, and $\SI{7.0}{\AA}$ using the pretraining configuration, but extended to 500 epochs with an additional learning rate reduction by $0.5$ at epoch 200. 

The training time of FCL for one epoch depends on the interval $[r_\text{min}, r_\text{max}]$ as this influences the average number of pairs per atom. The average number of pairs per atom during training is given by the expectation value of the cutoff radii, in our case $\mathbb{E}[\mathcal{U}(3.5,7.0)]=5.25$ (in \SI{}{\AA}). In agreement with this consideration, we found the time for one epoch of FCL to be comparable to training the baseline model with a static cutoff of $\SI{5.0}{\AA}$. Given the complexity of sampling sufficiently many per-atom radii $\rcut^{(i)}$, more epochs are required compared to static training. In this study, FCL models were trained for a maximum of 700 epochs in total (including 200 pretraining epochs), while static baseline models were trained for up to 500 epochs. 

\paragraph*{Cutoff Optimization}
As described in \autoref{subsec:optimization}, the cutoff radii can be optimized to improve the accuracy-cost tradeoff on calibration data. We optimize a FCL model on four subsets of its training data, each representing a distinct domain:

\begin{itemize}
    \item MC3D: 3D periodic inorganic crystals (bulk)
    \item MC2D: 2D periodic crystals (layers)
    \item SHIFTML-molcrys: 3D periodic molecular crystals
    \item SHIFTML-molfrags: finite neutral molecular fragments (no periodicity)
\end{itemize}

For each subset, we optimize the target function \eqref{eq:target_function_opt} with respect to element-wise cutoff radii $\Rcut_E$ using the Adam optimizer. We test five values of the tradeoff parameter: $\lambda \in \{10^{-2}, 10^{-3}, 10^{-4}, 10^{-5}, 10^{-6} \}$. Optimization runs for 10 epochs with learning rate $0.003$ and batch size 20, initializing all cutoff radii to $\SI{6.0}{\AA}$ while keeping model weights fixed. After optimization, we evaluate the resulting cutoff configurations on the corresponding test set subsets.

\begin{figure}[htb]
    \centering
    \includegraphics[width=0.75\linewidth]{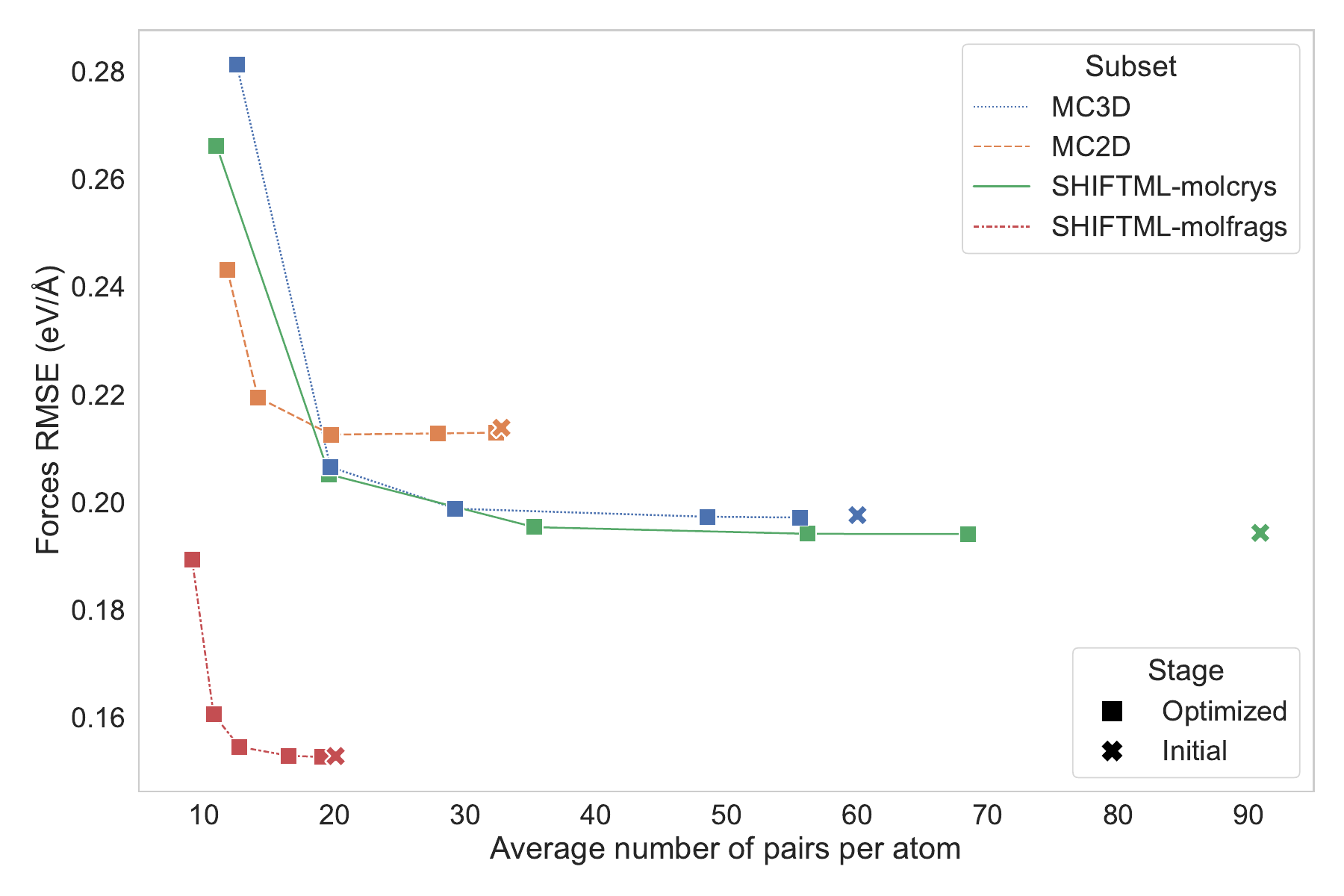}
    \caption{Force RMSE versus average number of pairs per atom for different MAD test subsets. Each curve shows results for varying $\lambda$ (from left to right: $10^{-2}, 10^{-3}, 10^{-4}, 10^{-5}, 10^{-6}$), demonstrating the accuracy-cost tradeoff achievable through cutoff optimization.}
    \label{fig:optimization_for_subsets}
\end{figure}

\autoref{fig:optimization_for_subsets} shows force RMSE on the test subsets as a function of computational cost (average pairs per atom) for optimized cutoff configurations. The subsets exhibit distinct error-cost profiles, with each $\lambda$ value yielding a Pareto-optimal configuration. The greatest accuracy changes occur for $\lambda \in [10^{-4}, 10^{-2}]$, while smaller $\lambda$ values ($10^{-6}$ to $10^{-4}$) produce minimal error increases despite significant cost reductions.

For the MC3D subset, reducing $\lambda$ from $10^{-6}$ to $10^{-4}$ decreases the average neighbor count from 54.4 to 29.3 (46\% reduction) while increasing force RMSE by only $0.83\%$ (from $\SI{197.21}{\meV\per\angstrom}$ to $\SI{198.85}{\meV\per\angstrom}$). Further reducing $\lambda$ to $10^{-3}$ yields 19.7 pairs per atom but raises the error to $\SI{206.58}{\meV\per\angstrom}$ ($3.89\%$ increase relative to the results obtained with $\lambda=10^{-4}$).

The absolute optimization potential varies across subsets. For MC2D and SHIFTML-molfrags, the initial uniform cutoff of $\SI{6.0}{\AA}$ is already close to the optimized configurations at $\lambda=10^{-6}$. A notable reduction in computational cost for both subsets can be achieved by increasing the cost sensitivity to $\lambda=10^{-4}$. In that case, the average number of pairs per atom measured for the MC2D dataset can be reduced by almost 40\% (32.8 to 19.7) and the test error even improves from \SI{213.87}{\meV \per \AA} to \SI{212.59}{\meV \per \AA}. On the subset with small molecule fragments (SHIFTML-molfrags), the computational cost can be reduced by almost 37\% while accuracy is degraded by $1.1\%$ ($\SI{152.92}{\meV \per \AA}$ to $\SI{154.62}{\meV \per \AA}$).

A greater optimization potential can be observed for the subset with molecular crystals (SHIFTML-molcrys): the initial configuration requires over 90 pairs per atom (RMSE: $\SI{194.36}{\meV\per\angstrom}$), while $\lambda=10^{-4}$ reduces this to 35 pairs per atom with only $0.54\%$ error increase ($\SI{195.42}{\meV\per\angstrom}$), that is, a reduction in computational cost by more than 60\%.

\begin{figure}[htb]
    \centering
    \includegraphics[width=0.75\linewidth]{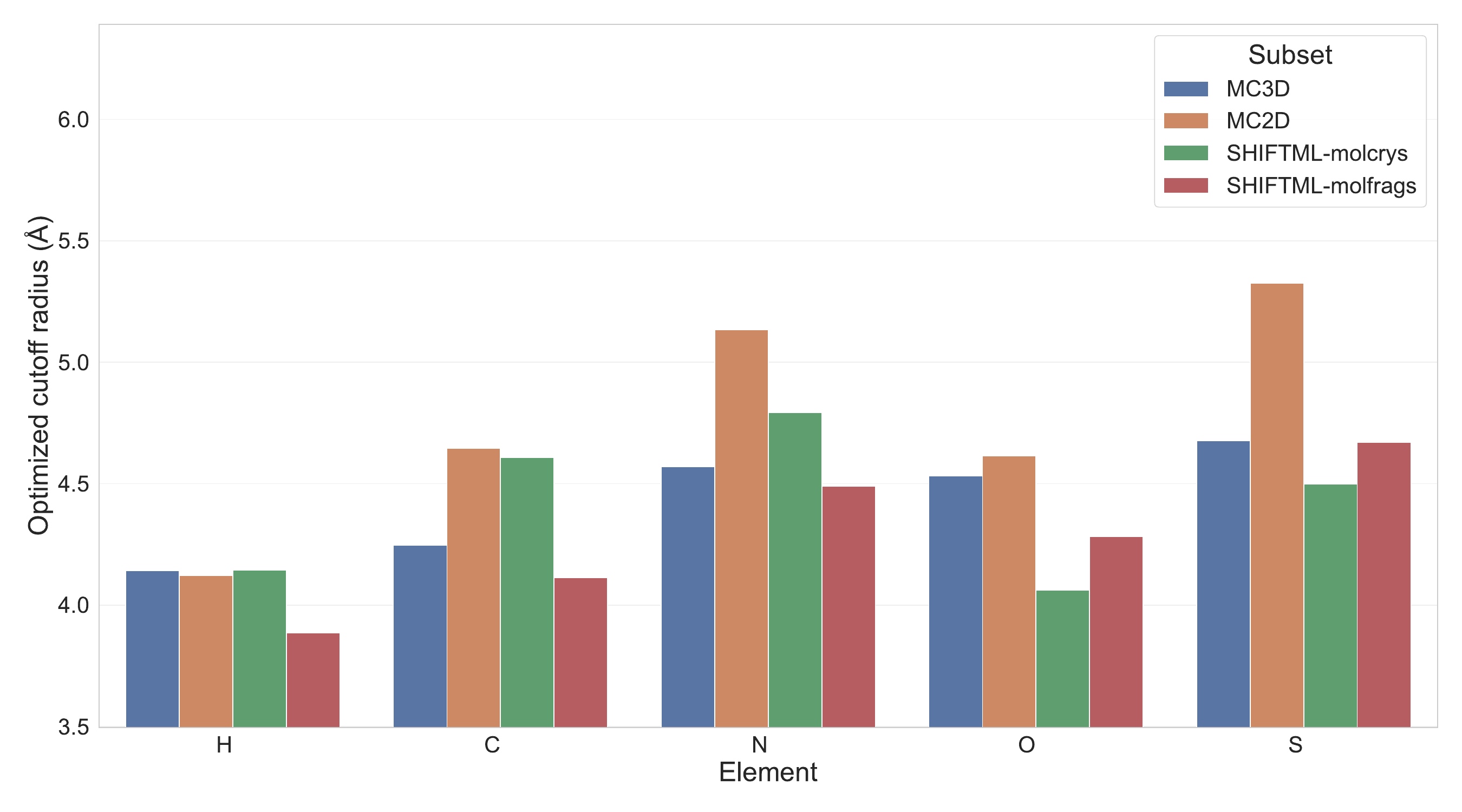}
    \caption{Optimized element-wise cutoff radii for $\lambda=10^{-4}$ across MAD subsets. Only elements present in all four subsets are shown.}
    \label{fig:element_cutoff}
\end{figure}

The optimized cutoff radii differ substantially across subsets, yielding domain-specific model configurations. \autoref{fig:element_cutoff} compares cutoffs for the five elements common to all subsets (\ce{H}, \ce{C}, \ce{N}, \ce{O}, \ce{S}). MC2D consistently produces the largest cutoffs, with sulfur reaching $\SI{5.4}{\AA}$—more than $\SI{0.5}{\AA}$ larger than in other subsets. 
Conversely, SHIFTML-molfrags yields the smallest cutoffs, particularly for \ce{H}, \ce{C}, and \ce{O}, consistent with the limited spatial extent of molecular fragments in this dataset.

\begin{figure*}
\centering
  \begin{subfigure}{0.24\textwidth}
    \includegraphics[width=\linewidth]{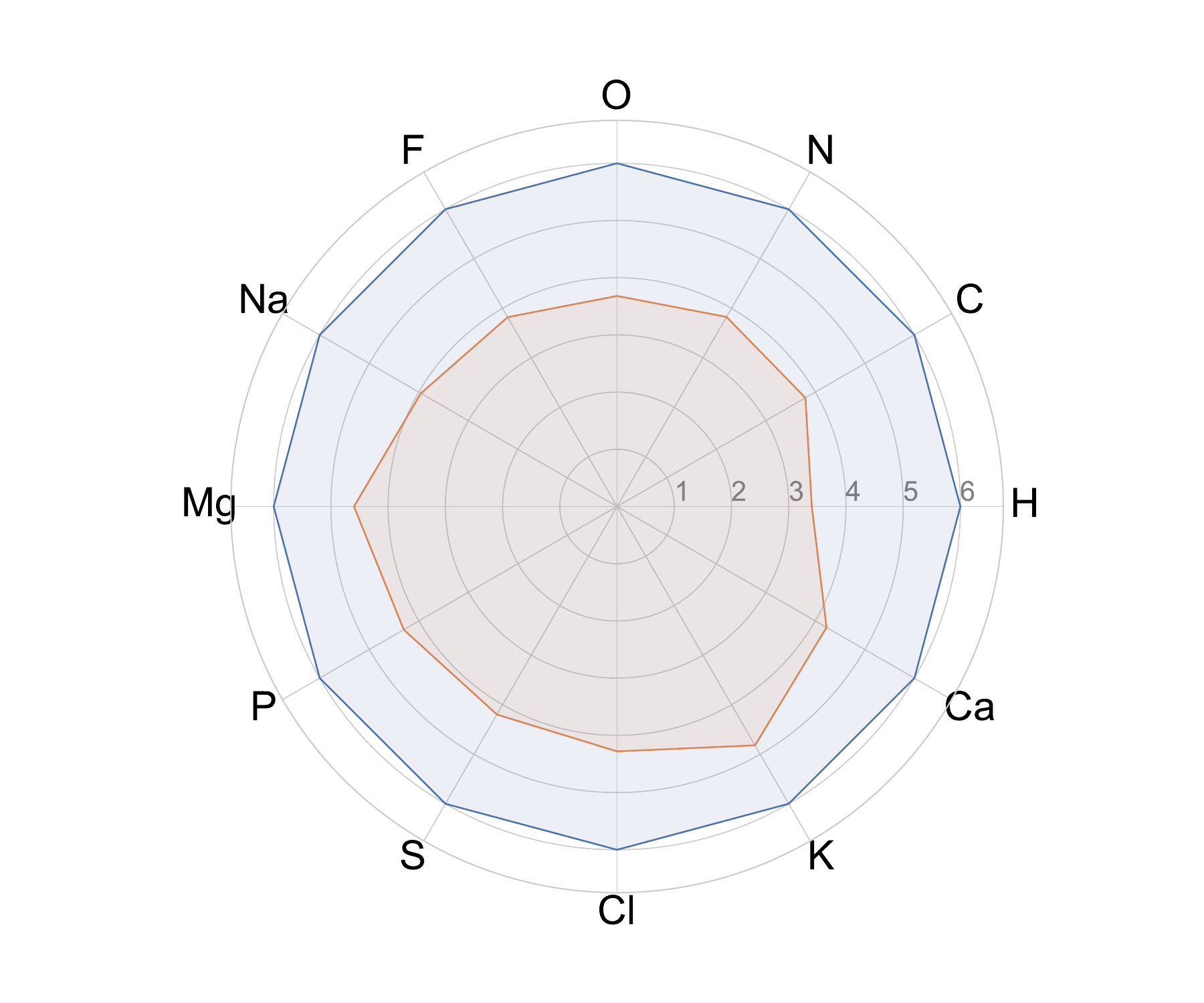}
    \caption{$\lambda=10^{-3}$}
    \label{fig:radarplots_a}
  \end{subfigure}%
  \hfill
  \begin{subfigure}{0.24\textwidth}
    \includegraphics[width=\linewidth]{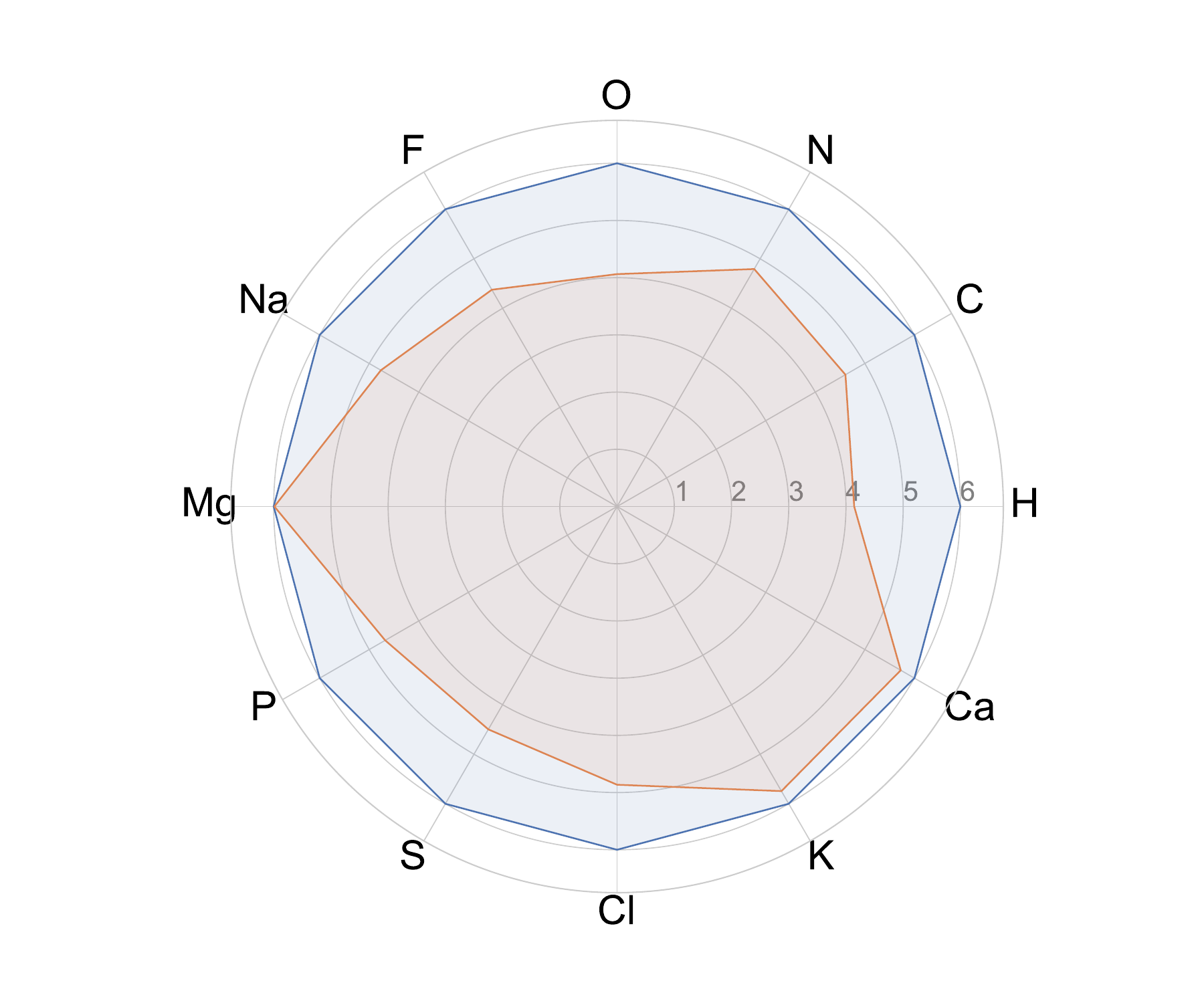}
    \caption{$\lambda=10^{-4}$}
    \label{fig:radarplots_b}
  \end{subfigure}%
  \hfill
  \begin{subfigure}{0.24\textwidth}
    \includegraphics[width=\linewidth]{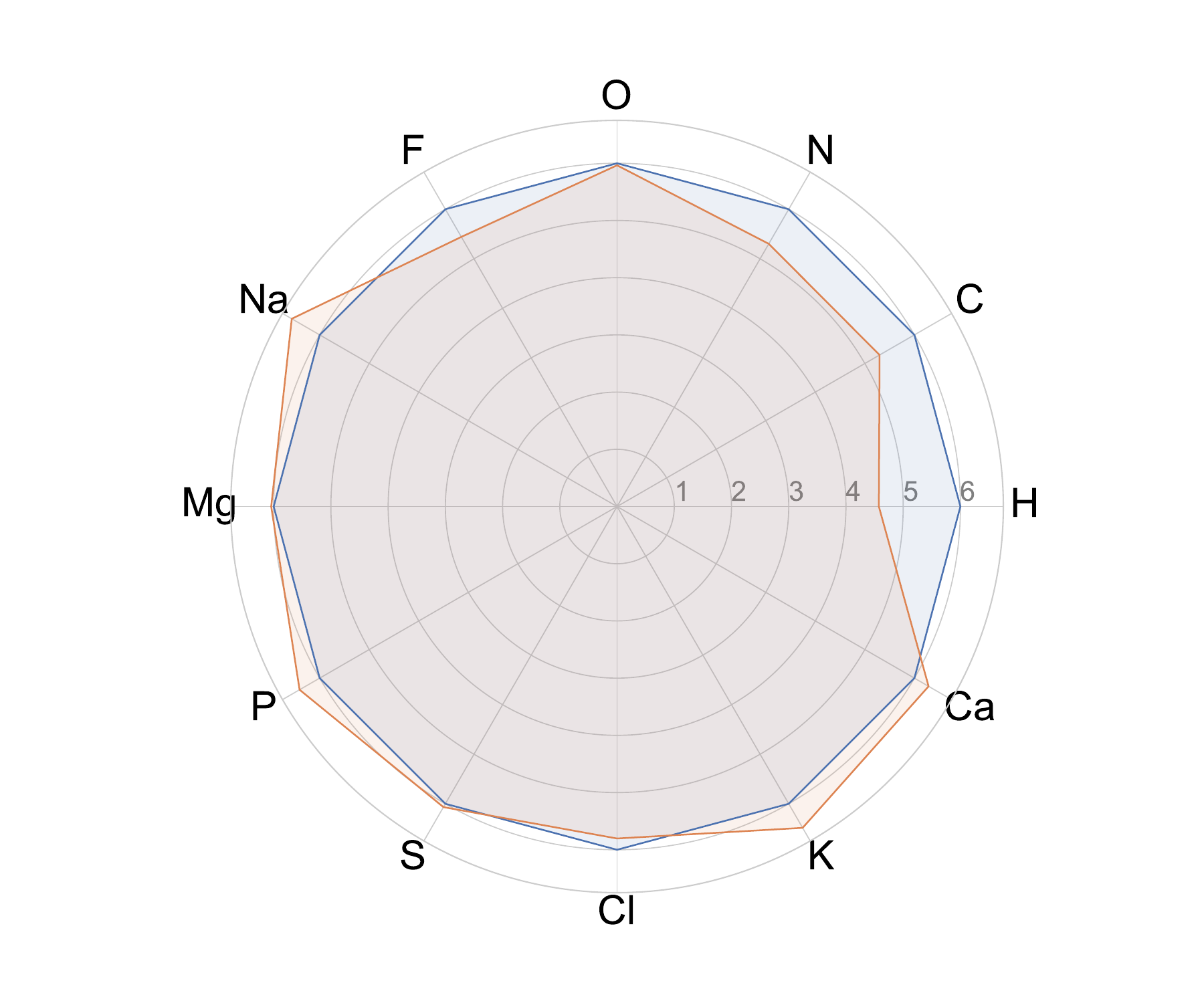}
    \caption{$\lambda=10^{-5}$}
    \label{fig:radarplots_c}
  \end{subfigure}
  \begin{subfigure}{0.24\textwidth}
    \includegraphics[width=\linewidth]{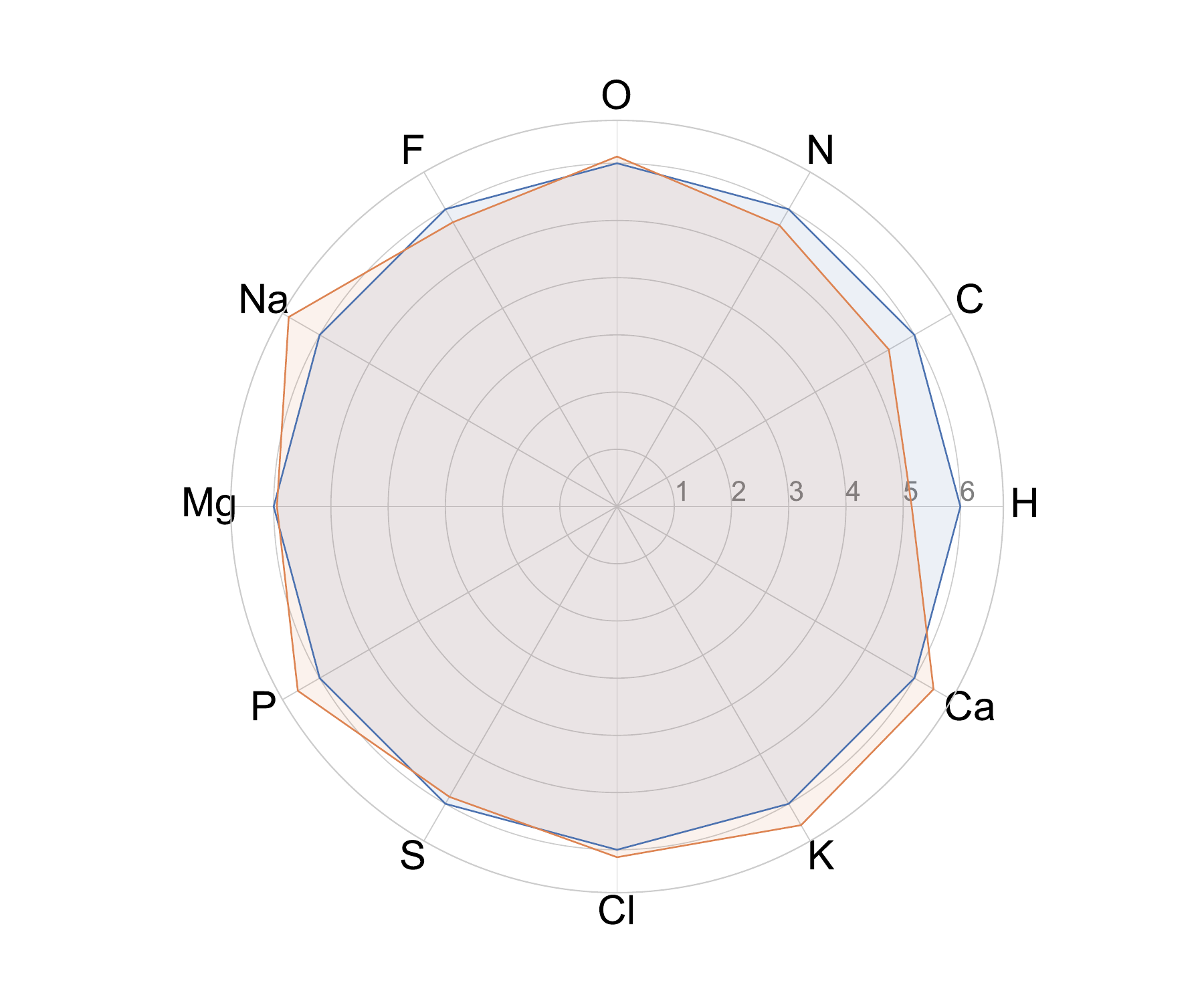}
    \caption{$\lambda=10^{-6}$}
    \label{fig:radarplots_d}
  \end{subfigure}
  \caption{Optimized cutoff radii (orange) versus initial values (blue) for SHIFTML-molcrys at different $\lambda$ values. Higher $\lambda$ emphasizes computational cost, producing smaller cutoffs across elements.}
  \label{fig:radarplots}
\end{figure*}

The tradeoff parameter $\lambda$ controls the position along the accuracy-cost Pareto front. \autoref{fig:radarplots} shows how $\lambda$ affects element-wise cutoffs for SHIFTML-molcrys. At high cost sensitivity ($\lambda=10^{-3}$), most cutoffs concentrate between $\SI{3.5}{\AA}$ and $\SI{4.5}{\AA}$. Light elements (\ce{H}, \ce{C}, \ce{N}, \ce{O}, \ce{F}) optimize to cutoffs below $\SI{4.0}{\AA}$, while heavier elements (\ce{Mg}, \ce{K}) maintain cutoffs greater than $\SI{4.5}{\AA}$. As $\lambda$ decreases to $10^{-6}$ (\autoref{fig:radarplots_d}), cutoffs diverge: light elements take values between $\SI{5.0}{}-\SI{6.0}{\AA}$, while alkali and alkaline earth metals (\ce{Na}, \ce{K}, \ce{Ca}) expand to $\SI{6.5}{}-\SI{7.0}{\AA}$. Despite a selective expansion for the latter elements, the average pair count reduces from 90 to approximately 70 (22\% decrease) compared to the uniform $\SI{6.0}{\AA}$ initialization.

Element abundance influences the optimization outcome. Since computational cost in \eqref{eq:target_function_opt} is affected by the element frequency, the optimization process tends to assign smaller cutoffs to more abundant elements at high $\lambda$ (see element frequencies in \autoref{tab:element_frequencies}). However, this trend is not absolute: at $\lambda=10^{-3}$, \ce{Na} receives a smaller cutoff than \ce{P}, \ce{S}, and \ce{Cl} despite being less abundant, indicating that the optimizer balances frequency against element-specific accuracy requirements.

\section{Discussion}
\label{sec:discussion}

Current foundational MLIPs rely on fixed cutoff radii that cannot be adjusted after training. While practitioners typically choose conservatively large cutoffs to ensure reliability across diverse systems, many specialized applications could achieve comparable accuracy with smaller cutoffs and reduced computational cost. We proposed Flexible Cutoff Learning to address this limitation by training MLIPs that remain flexible after training, enabling post-training optimization of the accuracy-cost tradeoff.

FCL relies on two key innovations. First, it treats the cutoff radius as a model input rather than a fixed architectural parameter, allowing the neural network to condition its predictions on cutoff values. Second, it assigns separate cutoff radii to each atom rather than using a single global cutoff, enabling fine-grained optimization. During training, atom-wise cutoff radii are randomly sampled from $\mathcal{U}(r_\text{min}, r_\text{max})$, exposing the model to diverse cutoff configurations. Interaction graphs for message passing are constructed using an arithmetic mean distance criterion based on per-atom cutoffs.

We demonstrated FCL by training a modified MACE architecture on the MAD dataset using a two-stage protocol: pretraining with a fixed $\SI{6.0}{\AA}$ cutoff, followed by FCL training with $r_\text{min}=\SI{3.5}{\AA}$ and $r_\text{max}=\SI{7.0}{\AA}$. The resulting model exhibits smooth accuracy-cost behavior across cutoff ranges, though performance degrades when evaluated near the upper training boundary at $r_\text{max}$. Analysis of diatomic energy curves revealed oscillatory behavior at this boundary, attributed to the lack of training examples beyond $r_{\text{max}}$ to regularize the model's behavior. For applications requiring cutoffs exceeding $\SI{6.5}{\AA}$, we recommend training with larger $r_{\text{max}}$ values.

The post-training flexibility enables gradient-based optimization of cutoff radii using a differentiable cost model. We defined a target function balancing empirical loss against computational cost through a tradeoff parameter $\lambda$, and optimized element-wise cutoffs on four MAD subsets representing distinct domains. Optimization potential varied  across subsets: MC2D and SHIFTML-molfrags showed a lower potential for improvement over the initial $\SI{6.0}{\AA}$ configuration, while 3D-periodic systems (MC3D and SHIFTML-molcrys) demonstrated higher gains. For SHIFTML-molcrys with $\lambda=10^{-4}$, optimization reduced the average pair count by more than 60\% (from 90 to 35 pairs per atom) while increasing test error by only 0.54\%. For MC3D with $\lambda=10^{-4}$, a 46\% cost reduction (54.4 to 29.3 pairs per atom) increased force RMSE by just 0.83\%. These results demonstrate that FCL enables substantial computational savings with minimal accuracy loss for selected subsets, though gains depend on the target domain.

The optimization process revealed systematic patterns in element-wise cutoffs. Subsets exhibited distinct cutoff profiles reflecting their interaction characteristics: MC2D almost consistently produced large cutoffs across elements, while SHIFTML-molfrags yielded small cutoffs consistent with limited molecular fragment sizes. Element abundance influenced optimization outcomes, with the cost term driving smaller cutoffs for frequent elements at high $\lambda$, though element-specific accuracy requirements modulated this trend. The tradeoff parameter $\lambda$ enabled traversal of the accuracy-cost Pareto front, with values between $10^{-6}$ and $10^{-4}$ providing substantial cost reductions while maintaining accuracy within 1\% of the initial configuration.

FCL complements existing finetuning workflows by optimizing the error-cost balance rather than accuracy alone. It enables a single foundational MLIP to be adapted to diverse applications through post-training cutoff optimization. While the reported cost reductions represent averages over entire subsets, individual applications may exhibit greater or lesser optimization potential depending on their specific structural characteristics.

\paragraph*{Limitations}

Our evaluation is limited in scope. We tested FCL on a single architecture (MACE) and dataset (MAD), leaving open questions about generalization to other MLIP architectures and chemical domains. Despite being model-agnostic from a theoretical perspective, broader validation across architectures and datasets is needed to establish FCL as a general training paradigm.

We focused on statistical error metrics (force RMSE) for evaluation. While such metrics are necessary for assessing model quality, they are not sufficient: models with low test errors can still exhibit unphysical behavior in molecular dynamics simulations or other tasks \cite{chiang2025}. Comprehensive validation would require testing FCL models in downstream applications such as structure optimization, phonon calculations, or MD trajectory stability. Whether optimized cutoff configurations maintain physical reliability across these tasks remains an open question.

We used the average number of pairs per atom as a computational cost proxy. This choice is appropriate for two-body MLIPs where memory and computation scale linearly with pair count. For three-body models (e.g., M3GNet, CHGNet), triplet count would be more suitable, and the optimization framework would require extension. We avoided measuring wall-clock time due to hardware-dependent effects: on GPUs, small systems may be dominated by I/O latency rather than computation, obscuring the theoretical linear scaling until hardware saturation is reached. The pair-count proxy eliminates such artifacts.

\section{Conclusion}
\label{sec:conclusion}

This work introduced \textit{Flexible Cutoff Learning} as an approach to train MLIPs that can be adapted to specific applications through gradient-based optimization of per-atom cutoff radii. To demonstrate FCL, we trained a modified MACE model on the MAD dataset by stochastically sampling atom-wise cutoff radii during training. We showed that the resulting model enables post-training optimization of the accuracy-cost tradeoff by adjusting cutoff radii. In particular, gradient-based optimization targeting this tradeoff reduced the average number of neighbor pairs per atom by more than 60\% on a molecular crystal subset while increasing the force error by less than 1\%. Looking forward, we believe that FCL represents a step toward general-purpose yet efficient MLIPs that can be cost-optimized for specific target applications without sacrificing the benefits of broad training.

\paragraph*{Competing interests}
The authors declare no conflict of interest.

\paragraph*{Code and Data availability}
The code for evaluating FCL models and performing cutoff optimization will be available after acceptance at \url{ https://github.com/Fraunhofer-SCAI/FlexibleCutoffLearning}. 

\paragraph*{Acknowledgments} This work was supported by the Fraunhofer Internal Programs under Grant No. PREPARE 40-08394. Furthermore, RO acknowledges funding by the DFG in the framework of the CRC 1639 ``NuMeriQS'' -- project no. 511713970. Furthermore, the authors thank Gerrit Schmieden for helpful discussions and valuable feedback. To improve readability, Claude Sonnet 4.5 R was used to improve the phrasing of certain parts of the text. No semantic changes were made by the model, and the authors take full responsibility for the content. 

\appendix

\printbibliography 

\newpage

\section{Adapting MACE to Flexible Cutoff Learning}
\label{subsec:FCL-mace}

In this work, we investigate the role of a flexible cutoff radius on the cost-error behavior of the model. To that end, we need to inform the model on the numerical value of the flexible cutoff radius, as described in \autoref{subsec:flexible_cutoff_learning}.   The key change is that the cutoff differs between edges and is computed from per-node cutoffs before the radial basis is applied. 

\paragraph*{Radial embedding} The MACE architecture employs radial functions $R^{(t)}_{k l_1 l_2 l_3}(r_{ij})$ at the $t$-th interaction block that modulate the equivariant tensor product for atoms $i$ and $j$ based on their distance $r_{ij}$. It is obtained by applying a Multi-Layer Perceptron $u$ to edge features $\mathbf{\Phi}_{ij}$ derived from the concatenation of $N_\text{RBF}$ radial basis functions $b_a(\cdot)$

\begin{equation}
    \mathbf{\Phi}_{ij} = [\phi_{1}(r_{ij}), \ldots, \phi_{N_\text{RBF}}(r_{ij})],
\end{equation}

where $\phi_{a}(r_{ij}) = b_{a}(r_{ij}) \cdot s(r_{ij}, r_{cut})$ and $s$ is the polynomial cutoff function \eqref{eq:polynomial_cutoff}. 

We augment the edge features of the original architecture as follows. We mix the per-node cutoff radii $\rcut^{(i)}$ into a per-edge cutoff $m_{ij}$ via the same mixing rule that is used for graph construction:
\[
m_{ij} = \mu(\rcut^{(i)}, \rcut^{(j)}).
\]

Edges are filtered by $r_{ij} \le m_{ij}$, and the flexible cutoff taper uses $m_{ij}$ rather than a global $\rcut$.

\begin{equation}
    \mathbf{\Phi}^{\mathrm{dyn}}_{ij} = u\big([\mathbf{b}(r_{ij}),\, m_{ij}]\big) \cdot s(r_{ij}, m_{ij}).
\end{equation}

The two-layer multilayer perceptron $u$ that maps the input features from dimension $N_\text{RBF} + 1$ to dimension $N_\text{RBF}$. Specifically, the network applies a linear transformation to 128 hidden units, followed by a SiLU (Sigmoid Linear Unit) activation function, and a final linear projection to the output dimension. Formally, the network can be expressed as:

\begin{equation}
u(\mathbf{x}) = \mathbf{W}_2 \cdot \text{SiLU}(\mathbf{W}_1\mathbf{x} + \mathbf{b}_1) + \mathbf{b}_2
\label{eq:radial_mlp}
\end{equation}

where $\mathbf{W}_1 \in \mathbb{R}^{128 \times (N_\text{RBF}+1)}$, $\mathbf{W}_2 \in \mathbb{R}^{10 \times 128}$, and $\mathbf{b}_1$, $\mathbf{b}_2$ are the corresponding bias vectors. In this work, we use $N_\text{RBF}=10$ (\texttt{num\_bessel=10}).

\paragraph*{Node features} We include information about the flexible cutoff radii to the node features of the MACE architecture. In particular, an additive contribution $\mathbf{q}(\rcut)$ is added to the initial node features $\mathbf{h}_i^{(0)}$ to obtain cutoff-informed initial node features $\tilde{\mathbf{h}}_i^{(0)}$ that enable the model to make its prediction adaptive to flexible cutoff radii:

\begin{equation}
    \tilde{\mathbf{h}}_i^{(0)} = \mathbf{h}_i^{(0)} + \mathbf{q}(\rcut^{(i)})
\end{equation}

The feature embedding of the cutoff radius $\mathbf{q}(\rcut^{(i)})$ is learned by a two-layer MLP $e$ similar to \eqref{eq:radial_mlp}, but with a hidden dimension of 64 rather than 128. 

\paragraph*{Hyperparameters}

Baseline models and modified MACE models use the hyperparameters provided in \autoref{tab:mace_hyperparams} (all other parameters correspond to default settings for \texttt{mace\_torch=0.3.14}).

\begin{table}[h]
\centering
\caption{MACE model hyperparameters.}
\label{tab:mace_hyperparams}
\small
\begin{tabular}{ll}
\toprule
\textbf{Parameter} & \textbf{Value} \\
\midrule
Cutoff radius ($r_{\text{max}}$) & variable \\
Bessel functions & 10 \\
Polynomial cutoff order & 5 \\
Correlation order & 1 \\
Maximum angular momentum ($\ell_{\text{max}}$) & 1 \\
Interaction layers & 2 \\
First interaction block & RealAgnosticInteractionBlock \\
Residual interaction block & RealAgnosticResidualInteractionBlock \\
Hidden irreps & $32 \times 0^e + 32 \times 1^o$ \\
MLP irreps & $16 \times 0^e$ \\
Gate activation & SiLU \\
Average neighbors & variable \\
Pair repulsion & False \\
Apply cutoff & True \\
Atomic interaction scale/shift & 1.0 / 0.0 \\
Number of elements & 100 ($Z = 1$--100) \\
Atomic energies & $0.0$ for all \\
\bottomrule
\end{tabular}
\end{table}

\begin{table}[h]
\centering
\caption{Hyperparameters for the static MACE models with different cutoff radii. The value for \texttt{avg\_num\_neighbors} is computed for the training set of the MAD dataset.}
\label{tab:rmax_avg_num_neighbors_mace}
\begin{tabular}{cc}
\toprule
\texttt{r\_max} & \texttt{avg\_num\_neighbors} \\
\midrule
4.0 & 19.3 \\
5.0 & 38.2 \\
6.0 & 65.4 \\
7.0 & 104.1 \\
\bottomrule
\end{tabular}
\end{table}

Note that the cutoff radius of the MACE architecture \texttt{r\_max} should not be confused with $r_\text{max}$ in the main text, which denotes the upper bound for sampling flexible cutoff radii during training. \texttt{avg\_num\_neighbors} is used for normalization of messages in the MACE architecture and depends on \texttt{r\_max}. The corresponding values for the static models are provided in \autoref{tab:rmax_avg_num_neighbors_mace}. For the FCL models, we used a value of 50 throughout all experiments.

\section{MAD SHIFTML-molcrys frequency}

The element frequencies for the SHIFTML-molcrys subset of the MAD dataset can be found in \autoref{tab:element_frequencies}.

\begin{table}[h]
  \centering
  \begin{tabular}{c|c}
    Element & Fraction \\[0.15em]
    \hline
    H  & $4.20 \cdot 10^{-1}$ \\
    C  & $3.27 \cdot 10^{-1}$ \\
    N  & $1.05 \cdot 10^{-1}$ \\
    O  & $9.49 \cdot 10^{-2}$ \\
    F  & $1.77 \cdot 10^{-2}$ \\
    Na & $4.95 \cdot 10^{-4}$ \\
    Mg & $8.80 \cdot 10^{-5}$ \\
    P  & $4.38 \cdot 10^{-3}$ \\
    S  & $1.63 \cdot 10^{-2}$ \\
    Cl & $1.33 \cdot 10^{-2}$ \\
    K  & $4.42 \cdot 10^{-4}$ \\
    Ca & $2.71 \cdot 10^{-4}$ \\
    \hline
  \end{tabular}
  \caption{Element frequencies in SHIFTML-molcrys training data.}
  \label{tab:element_frequencies}
\end{table}

\section{Plots for Energy Predictions}

\begin{figure}[htbp]
  \centering
    \includegraphics[width=0.6\textwidth]{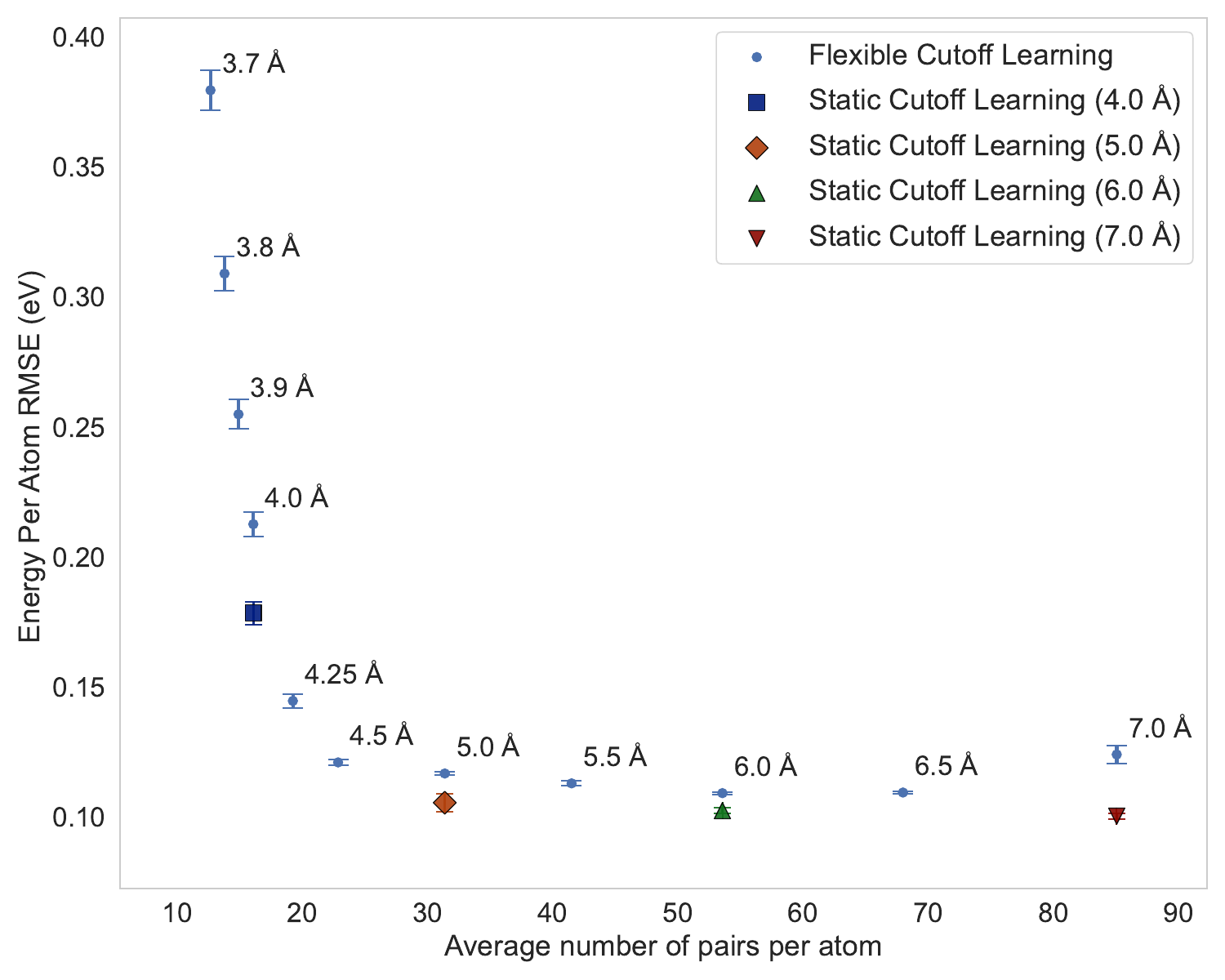}
  \caption{RMSE of energy per atom as a function of a global cutoff radius}
  \label{fig:Error-Cost-Curves-Test-Energy}
\end{figure}

\begin{figure}[htbp]
    \centering
    \includegraphics[width=0.6\linewidth]{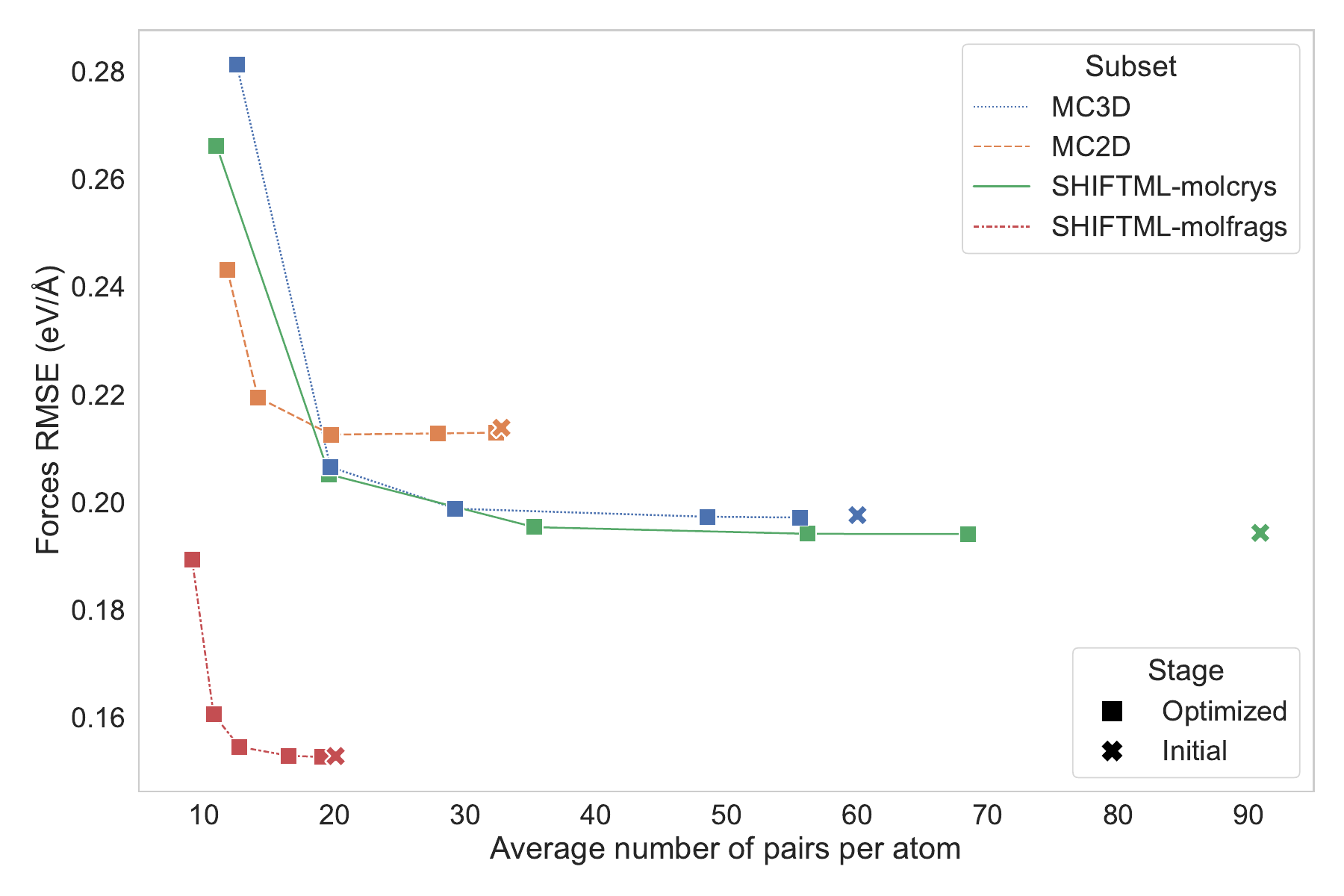}
    \caption{Optimizing the per-element cutoff radii on different subsets of the MAD dataset for different values of  from left to right) improves the error-cost ratio of a model. A single FCL-model is evaluated on different subsets from the MAD dataset of the test data using the element-wise cutoff radii $\Rcut_E$ from the optimization process with different values for $\lambda \in \{10^{-2}, 10^{-3}, 10^{-4}, 10^{-5}, 10^{-6} \}$ (from left to right). Shows the RMSE of the energy per atom.}
    \label{fig:optimization_for_subsets_energy}
\end{figure}

In the main text, we focus on force predictions of the model, as the forces are usually more relevant for applications such as molecular dynamics simulations or geometry optimizations. For completeness, the corresponding plots of the energy predictions are found in \autoref{fig:Error-Cost-Curves-Test-Energy} and \autoref{fig:optimization_for_subsets_energy}.

\section{Diatomic Curves Full Range}

\begin{figure*}
  \begin{subfigure}{0.33\textwidth}
    \includegraphics[width=\linewidth]{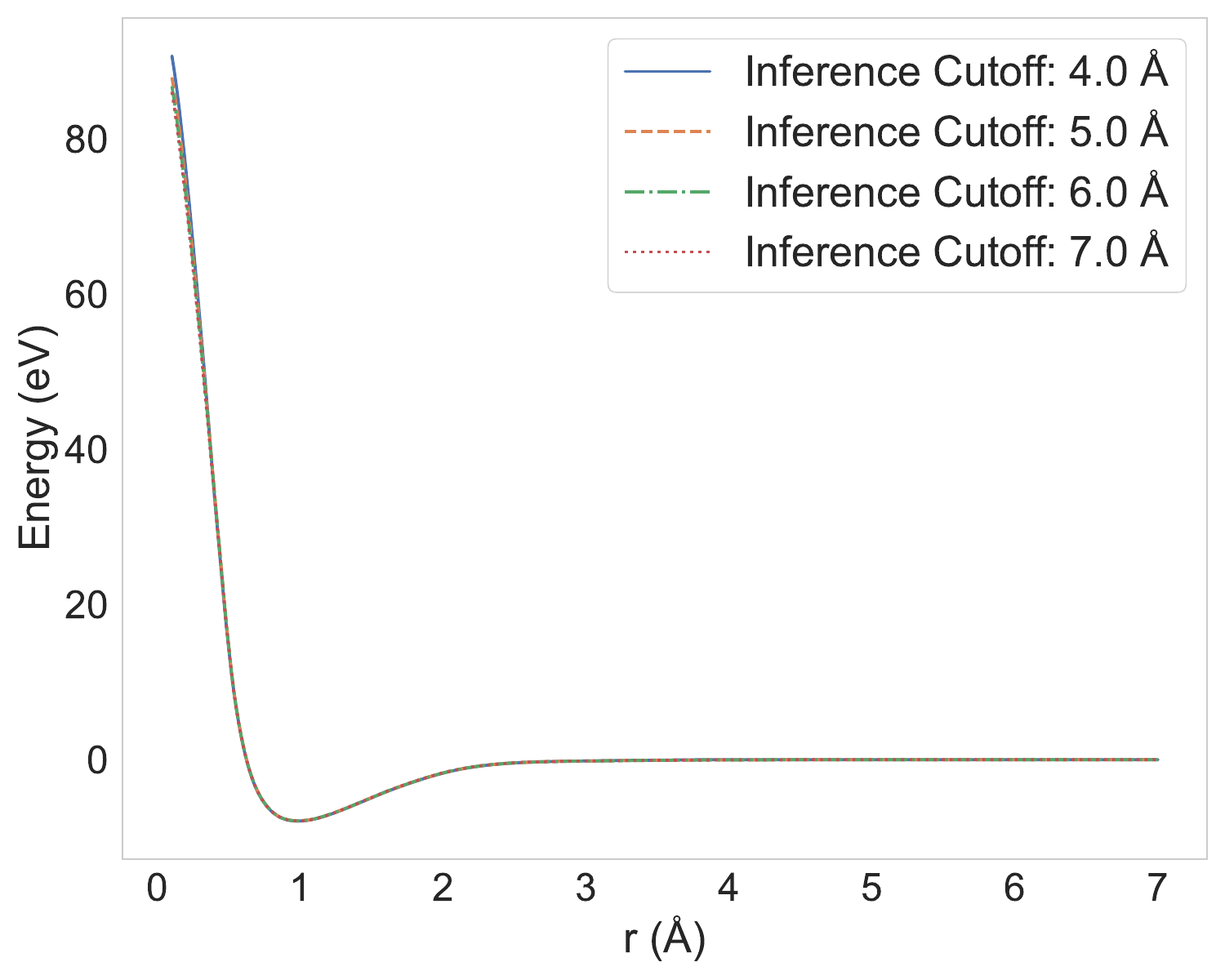}
    \caption{H-O}
  \end{subfigure}%
  \hfill
  \begin{subfigure}{0.33\textwidth}
     \includegraphics[width=\linewidth]{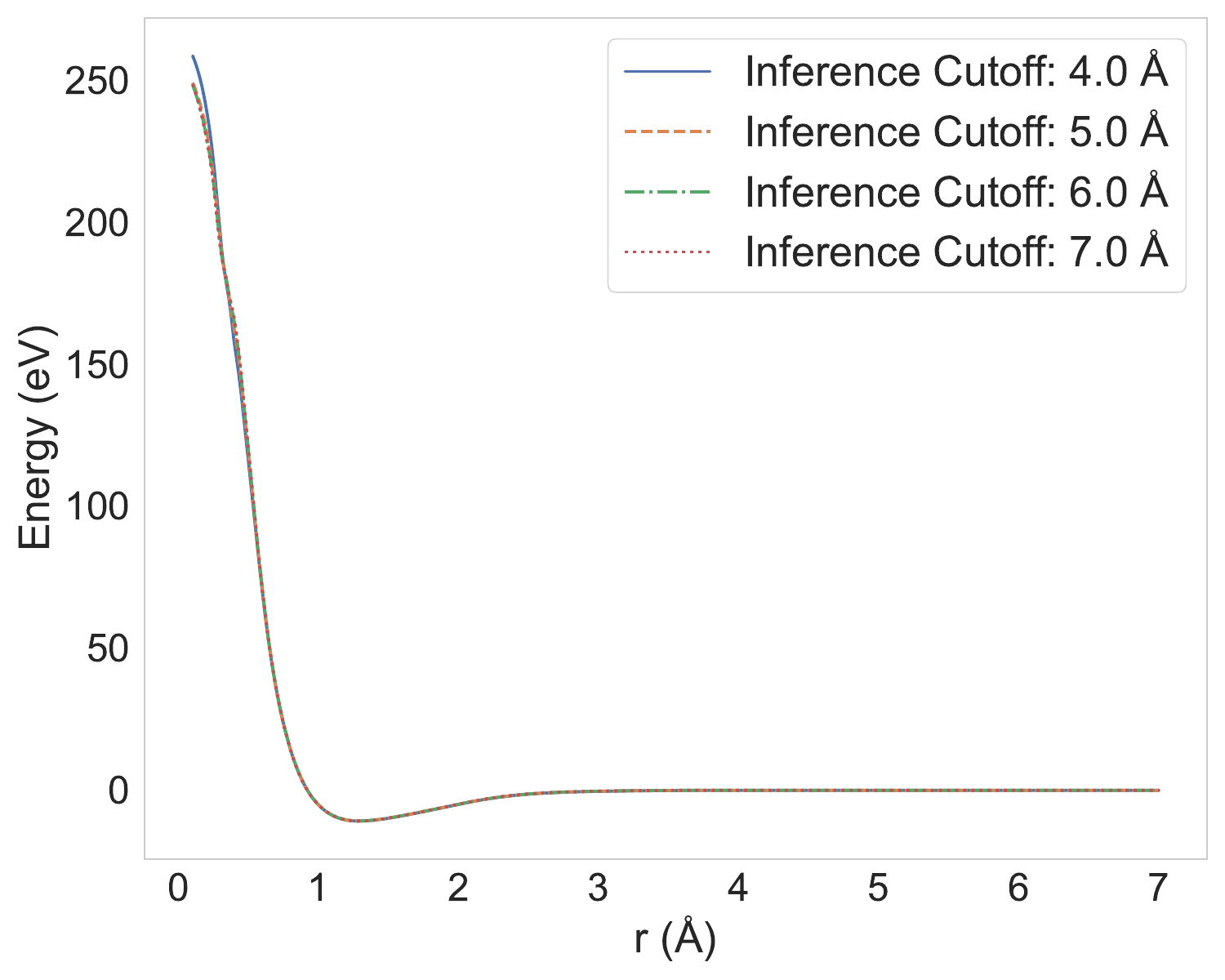}
    \caption{C-C}
  \end{subfigure}%
  \hfill
  \begin{subfigure}{0.33\textwidth}
    \includegraphics[width=\linewidth]{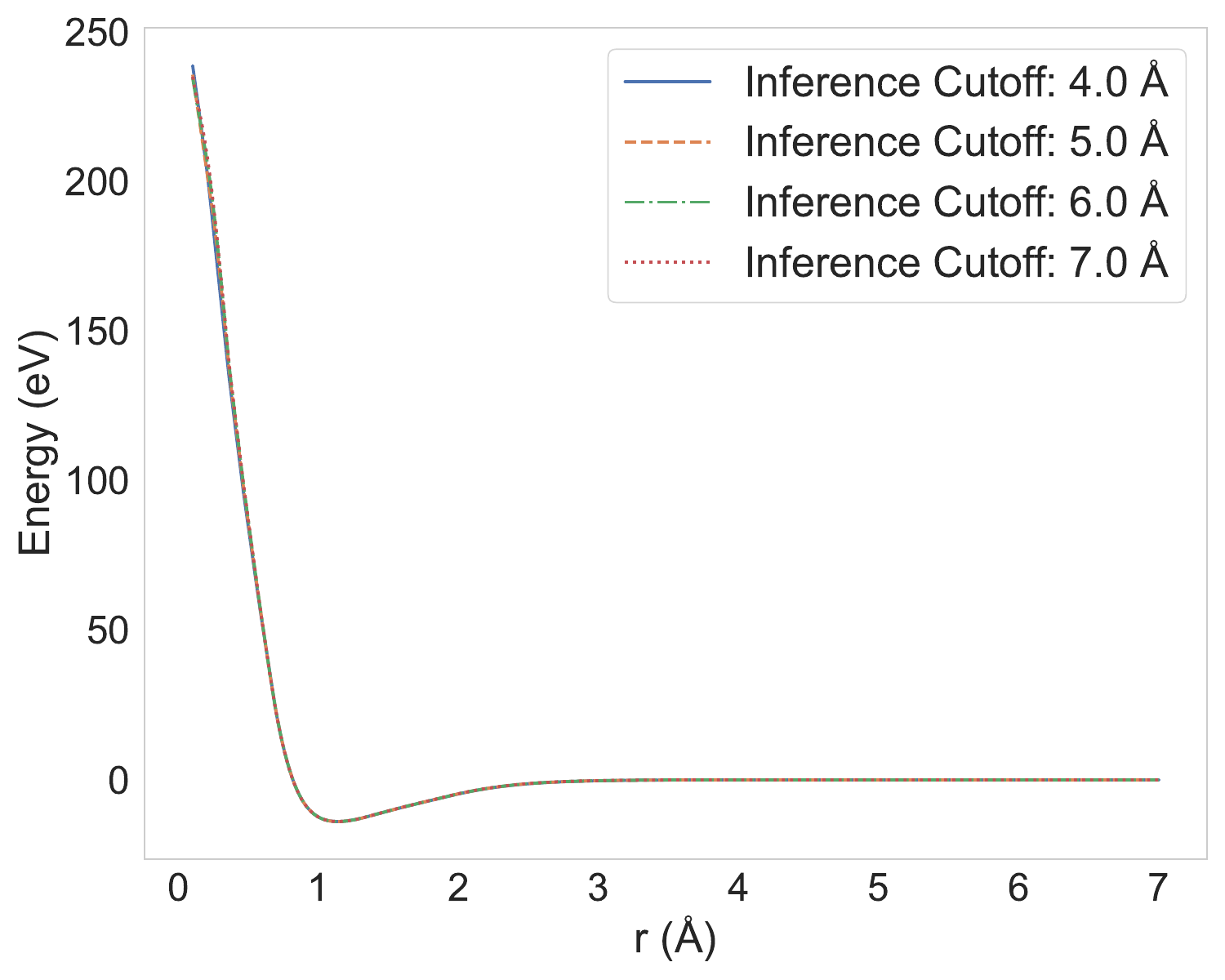}
    \caption{C-O}
  \end{subfigure}
  \caption{For the full range on the x-axis: Energy as a function of distance in diatomic systems for different values of the cutoff radius. All curves are obtained by evaluating a single model.}
  \label{fig:pair_curves_full}
\end{figure*}

\autoref{fig:pair_curves_full} illustrates the diatomic pair curves as shown in the main text over the full range of interatomic distances $r$. 

\end{document}